\documentclass[pre,twocolumn,superscriptaddress,showpacs]{revtex4}

\usepackage[dvips]{graphicx}
\usepackage{amsmath}
\usepackage{latexsym}

\begin{document}

\title{Topological defects in the  crystalline state of 
   one-component plasmas of non-uniform density}
\date{\today}

\author{A. Mughal and M. A. Moore}
\affiliation{School of Physics and Astronomy, University of 
Manchester, Manchester M13 9PL, U.K.}

\begin{abstract}
  We study the ground state properties of classical Coulomb charges
  interacting with a $1/r$ potential moving on a plane but confined
  either by a circular hard wall boundary or by a harmonic potential.
  The charge density in the continuum limit is determined analytically
  and is non-uniform.  Because of the non-uniform density there are
  both disclinations and dislocations present and their distribution
  across the system is calculated and shown to be in agreement with
  numerical studies of the ground state (or at least low-energy
  states) of $N$ charges, where values of $N$ up to 5000 have been
  studied.  A consequence of these defects is that although the
  charges locally form into a triangular lattice structure, the
  lattice lines acquire a marked curvature. A study is made of
  conformal crystals to illuminate the origin of this curvature.  The
  scaling of various terms which contribute to the overall energy of
  the system of charges viz, the continuum electrostatic energy,
  correlation energy, surface energy (and so on) as a function of the
  number of particles $N$ is determined. ``Magic number'' clusters are
  those at special values of $N$ whose energies take them below the
  energy estimated from the scaling forms and are identified with
  charge arrangements of high symmetry.
\end{abstract}

\pacs{61.72.Bb, 61.72.Mm, 61.72.Lk} 

\maketitle

\section{Introduction}

Classical charges moving on a plane and repelling each other via a
Coulomb $1/r$ potential have a ground state which is a triangular
lattice (for appropriately chosen periodic boundary conditions).  This
is a two-dimensional (2D) example of Wigner crystallization and is
known to occur in diverse areas of physics, such as  electrons trapped in surface 
states of liquid helium \cite{cole}, colloidal
suspensions and quantum dots.  In this paper we examine situations
when the charges do not have a uniform density across the system.
This occurs for example when the charges are confined by a circular hard wall or
when they are confined by a harmonic potential.

From a geometrical view point a perfect crystal lattice is a
periodically repeating arrangement of identical structural cells,
which fit together without gaps or overlap. The question we seek to
answer, for a classical Wigner crystal of non-uniform density, is: how
much of what we mean by ``crystal lattice'' still applies to the
resulting structure?  On the one hand, one expects the structure of
the lattice to be locally triangular; so that each lattice site has
six nearest neighbors, since this is the optimal energy arrangement
for uniform density. On the other hand, due to the changing density
not all the symmetries of the triangular space group, such as the
translational and rotational invariances can continue to apply.  We
seek to understand how this conflict is resolved.  We are particularly
interested in knowing if the resulting structure can be understood
within the framework of elasticity theory, and if so, what role is
played by plastic deformations, such as dislocations and
disclinations. In addition, using a continuum model, we shall attempt
to quantify the scaling with $N$ of various phenomena associated with
the structure of the cluster of charges, such as the energy,
correlation energy (see below for its definition), surface energy and
so on. This will in turn allow us to develop a link between symmetry
and the energy of the lattice; we expect states with a high degree of
symmetry to have a particularly low energy. Such states are known in
the literature as ``magic number'' states.

An experimental realization of a system of 2D charges confined by a
hard wall might be electrons trapped in surface states of liquid
$^4$He \cite{cole}.  The hard wall potential could be effected by a
circular boundary made from an electrical insulator. A harmonic
confining potential also produces a non-uniform density across the
system and has been studied  by Koulakov and Shklovskii, see
\cite{koulakov2} and \cite{koulakov}. Their study is relevant to the
properties of quantum dots.  It was found that the density of the
charges was greatest at the center and diminished, upon approaching
the edge (whereas with hard-wall confinement, the density rises from
the center of the disc towards the wall because of the repulsion
between the charges).  Their simulations showed that although the
charge density was not uniform, nevertheless the lattice was locally
triangular.  They showed that this is possible because the changing
lattice density is accompanied by plastic deformations.

It is a consequence of topology that both the hard-wall and the
harmonic potential systems must contain an excess of 6 positive
disclinations or pentagonal regions. Furthermore, the changing density
introduces disclinations throughout the cluster. Disclinations also
occur at the edges of the clusters to allow the lattice to adapt to
the imposed circular structure.  In addition to disclinations, the
cluster includes dislocations (a tightly bound five-seven coordinated
disclination pair). Unlike disclinations the total number of
dislocations is not fixed by topology; dislocations are present to
reduce the strain energy in the crystal which is induced by the
circular edge and the disclinations. Dislocations are present in large
numbers near the lattice edge, where they form a cloud around any
disclinations and help reduce the large elastic stress induced by the
latter. However, it was found by Koulakov and Shklovskii that
disclinations and dislocations are to be found also in the lattice
interior and act as a mechanism for reducing the strain energy there.

A number of other authors have carried out numerical simulations in
situations where the classical 2D Wigner crystal has a non-uniform
density.  Bedanov and Peeters \cite{bedanov} considered particles
interacting via the pure Coulomb potential and confined either by a
hard wall or a parabolic potential. Simulations on the hard wall
problem have also been carried out by Kong et al \cite{kong}; in
addition to the Coulomb interaction they also consider the dipole and
Yukawa potentials.  Ying-Ju Lin and Lin I \cite{ying} have studied a
number of systems with different interaction and confining potentials.
Our work confirms and extends these earlier numerical studies, but
also includes an account of our attempts to understand the numerical
results.

The paper is organized as follows. In Sec. II, we present the systems
we study and our numerical approach. In Sec. III we expand on the work
by Koulakov and Shklovskii and derive the continuum approximation for
the cluster of charges. We can then give an analytical calculation for
the density of charges in the continuum limit, and for the density of
dislocations and disclinations. We give a series expansion which we
believe describes the various contributions to the ground state energy
of the cluster. These results are compared to numerical experiments in
Sec. IV. The discussion is in Sec. V, especially of the striking
lattice curvature effect visible in our studies of large clusters
which is compared to that seen in conformal crystals.

\section{Numerical Approach}
The energy of a cluster of N charges confined to a disk of radius R,
by a hard-wall potential is given by
\begin{equation}
E^H
=
{\sum_i^N}V({\bf r}_i)
+
{\sum_{i<j}^N}\frac{1}{|{{\bf r}_i}-{{\bf r}_j}|}, 
\label{eq:totalenergy}
\end{equation}
where 
\begin{eqnarray}
V({\bf r}_i)=\left\{ \begin{array}{ll}
0      & \mbox{for ${r_i}<R$} \\
\infty & \mbox{for ${r_i}\geq R$}.
\end{array}
\right. 
\nonumber
\end{eqnarray}
The energy of a cluster of N charges in a parabolic confining
potential is given by
\begin{equation}
E^P
=
A{\sum_i^N}{r_i}^2
+
{\sum_{i<j}^N}\frac{1}{|{{\bf r}_i}-{{\bf r}_j}|}, 
\label{eq:totalenergy_koulakov}
\end{equation}
where we set A=1/2.

Finding the global minimum for a function such as
$E^H$ or $E^P$ is a very difficult task. The number of metastable
states proliferate exponentially with $N$; consequentially the global
minimum is obscured by a vast number of local minima with energies
close to that of the global minimum. There exist a number of heuristic
methods for such problems. Although there is no guarantee of finding
the global minimum, it is possible to find states close to it.

We found that for the hard-wall system the standard Metropolis
simulated annealing algorithm to be more effective than a conjugate
gradient algorithm \cite{press}. For a system with $N$ charges the
simulated annealing algorithm was run with typically
$N\times(5\times10^6)$ Monte Carlo steps. The temperature of the
simulation was decreased linearly. The average displacement of the
charges at each temperature step was chosen by an automatic process to
give an acceptance probability of $0.5 \pm 0.01$. Promising states
were reheated and annealed repeatedly to iron out as many defects as
possible. Finally the results were put through a conjugate gradient
algorithm to remove any residual strains.

For the harmonic (parabolic) potential case, we found the conjugate
gradient algorithm to be as effective as simulated annealing. We used
the former method for this system as it ran faster.  Results were
generated starting from an initial random configuration, which had a
radial density profile matching the continuum limit density given
below in Eq. ($\!\!$~\ref{eq:koulakov_density}).

\section{The Continuum Limit} 

In the following we develop the continuum model of the two systems.
For the case of parabolic confinement, many of the important results
have already been derived by Koulakov and Shklovskii, see
\cite{koulakov2} and \cite{koulakov}, so where appropriate we shall
simply quote the relevant result. The bulk of the material in this
section is concerned with the system with a hard-wall confining
potential.

In the following, for the system with the hard-wall confining
potential, we use a variational approach to derive the (non-uniform)
charge density in the continuum limit. Next we demonstrate that as a
consequence of the non-uniform density the system interior will
contain topological defects, where the density of these defects
depends on the rate of change of the density. Finally, for the system
with a hard-wall confining potential, a series is developed which
includes the contributions to the energy of the cluster which scale
smoothly with system size.

\subsection{Charge Density}

For charges confined by a hard-wall potential, the energy expression 
in Eq. (\ref{eq:totalenergy}) can be approximated by 
the integrals over the disc $r\le R$,
\begin{equation}
E
=
\frac{1}{2}
\int\!\,d^2{r}\int\!\,d^2{r}
\frac
{
\rho^{H}{({\bf r})}\rho^{H}{({\bf r'})}
}
{
|{\bf r}-{\bf r'}|
}
.
\label{eq:energy}
\end{equation}
The continuum 
approximation treats the density $\rho^{H}({\bf r})$ as a 
smooth function rather than the sum of delta functions
\begin{equation}
\rho^{H}({\bf r})= \sum_{i=1}^{N}\delta({\bf r}-{\bf r}_i),
\label{eq:rhodef}
\end{equation}
where ${\bf r}_i$ is the position of the $i^{th}$ charge.
One then minimizes the energy of the cluster, with respect to the smooth function 
$\rho^{H}{({\bf r})}$, subject to the constraint that the number of particles
\begin{equation}
N=\int\!\,d^2{r}\rho^{H}{({\bf r'})},
\label{eq:Nparticles}
\end{equation}
is constant.
Introducing the Lagrange multiplier $\mu$ the
constrained equation is
\[
E
=
\int
\rho^{H}{({\bf r'})}
\left[
\frac{1}{2}
\int\!\,d^2{r}
\frac
{\rho^{H}{({\bf r})}}
{|{\bf r}-{\bf r'}|}
-
\mu
\right]
d^2{r'}
.
\]
A variation in the energy is given by
\begin{equation}
\delta E = E{[\rho^{H}{({\bf r})} +\delta\rho^{H}{({\bf r})}]}-E{[\rho^{H}{({\bf r})}]},
\label{eq:evariation}
\end{equation}
where $\delta\rho^{H}{({\bf r})}$ represents a small change in the charge
density. Keeping only terms up to first order,
Eq. ($\!\!$~\ref{eq:evariation}) gives
\begin{equation}
\delta E
=
\int\!\,
\delta\rho^{H}{({\bf r'})}
\left[
\int\!\,d^2{r}
\frac
{\rho^{H}{({\bf r})}}
{|{\bf r}-{\bf r'}|}
- 
\mu
\right]
d^2{r'}
.
\end{equation}
To make the functional derivative stationary we require
\begin{equation}
\int\!\,
d^2{r}
\frac
{
\rho^{H}{({\bf r})}
}
{
|{\bf r}-{\bf r'}|
}
-
\mu
=
0.
\label{eq:n=1(2d)problem}
\end{equation}
To solve this integral equation it is convenient to write the integral
in Eq. ($\!\!$~\ref{eq:n=1(2d)problem}) in terms of radial and angular
variables:
\begin{eqnarray}
\mu
&=&
\int^{R}_{0}\!\,
\rho^{H}(r)rdr
\int_{0}^{2\pi}\!\,
\frac
{d\theta}
{(r^{2} +{r'}^{2} -2rr'\cos\theta)^ {\frac{1}{2}} }
\nonumber 
\\
&=&
\int^{R}_{0}\!\,
dr
\frac
{4r\rho^{H}(r)}
{r+r'}
K
\left(
\frac
{2\sqrt{rr'}}
{r+r'}
\right),
\nonumber
\end{eqnarray}
where $K(k)$ is an elliptical integral of the first kind. This is a
Fredholm integral equation of the first kind which has the radially
symmetric solution
\cite{polyanin}
\begin{equation}
\rho^{H}(r)
=\frac{\mu}{\pi^2}
\frac{1}{\sqrt{{R^2}-{{r}^2}}}.
\label{eq:eintegralEqSoln1}
\end{equation}
To determine the Lagrange multiplier we substitute Eq.
($\!\!$~\ref{eq:eintegralEqSoln1}) into Eq.
($\!\!$~\ref{eq:Nparticles}), which gives $\mu=N\pi/2R$, and we
finally have
\begin{equation}
\rho^{H}(r)
=
\frac
{N}
{2\pi R^2}
\frac
{1}
{\sqrt{ 1 - \left(\frac{r}{R}\right)^{2} } }.
\label{eq:density}
\end{equation}
Incidentally, this result is the density profile obtained if a
hemispherical shell of charge is projected onto a plane.  

It is important to note the physical meaning of the Lagrange
multiplier $\mu$; it is the electric potential at any point
inside the disk when the charge is distributed according to
Eq. ($\!\!$~\ref{eq:density}). This can be seen by referring back to
the original formulation of the problem as given by
Eq. ($\!\!$~\ref{eq:n=1(2d)problem}).

The number of charges within a distance $r$ from the center of the
disc is given by integrating Eq. ($\!\!$~\ref{eq:density}) over the
region $r^{\prime}\leq r$:
\begin{equation}
N^{H}(r)
=
N
\left(
1
-
\sqrt{
1
-
\left(
\frac{r}{R}
\right)^2
}
\right)
.
\label{eq:Ncharges}
\end{equation}

The charge density in the continuum limit for a cluster of charges in
a parabolic confining potential is \cite{koulakov}
\begin{equation}
\rho^{P}(r)=\rho_{o}\sqrt{1-\left(\frac{r}{R}\right)^2},
\label{eq:koulakov_density}
\end{equation}
where 
\[
R=\left(\frac{3\pi N}{8A}\right)^\frac{1}{3} \;\; \text{and} \;\; \rho_{o}=\frac{4AR}{\pi^{2}}.
\]
The total number of charges within a distance $r$ from the center of 
the disc  is 
\begin{equation}
N^{P}(r)
=
N
\left(
1
-
\left( 1- \left(\frac{r}{R}\right)^{2} \right)^{\frac{3}{2}}
\right).
\label{eq:KNcharges}
\end{equation}

\subsection{Density of Defects}

Volterra dislocations are plastic imperfections characteristic of a
deformed solid \cite{Chaikin}. For a 2D lattice the only relevant
Volterra dislocations are the edge dislocations and the wedge
disclination; these we discuss in turn below.

A dislocation in a perfect lattice can be created by making a cut in
the lattice, translating the cut edges with respect to each other and
inserting/removing material. This process can also be viewed as the
insertion/removal of a half plane of atoms, it is characterized by a
discrete Burgers vector {\bf B} which measures the amount by which the
Burgers circuit around the dislocation fails to close \cite{weertman}.

A disclination can be created in a perfect lattice by making a cut,
rotating the cut edges with respect to each other (this then defines a
disclination axis) and inserting/removing a wedge of material.  After
the wedge is inserted/removed the whole construction is welded
together and allowed to relax. The closure failure of a Burgers
circuit around a disclination is given by
\begin{equation}
{\bf B}={\bf \Omega} \times {\bf r}, 
\label{eq:closure_faliure_around_a_disclination}
\end{equation}
where {\bf r} is the distance from the disclination axis ${\bf
  \Omega}$, and ${|{\bf \Omega}|}$ is the wedge angle \cite{friedel}.
In a real lattice the wedge angle is quantized by the lattice
symmetry, thus in a triangular lattice matter is inserted into the
lattice if $|{\bf \Omega}|=+2\pi/6$ and removed if $|{\bf
  \Omega}|=-2\pi/6$, which corresponds to a heptagon or a pentagon in
the crystal lattice respectively.

It is important to know the relationship between dislocations and
disclinations. A dislocation is a tightly bound pair of disclinations
of the opposite sign. A disclination can be decomposed into a series
of dislocations, each of which have the same sign. For an illustration
of both these points see \cite{friedel}.

Disclinations can be present in the ground state of a system for two
reasons; either because they are demanded by topology (a consequence
of Euler's theorem), or because the lattice density is non-uniform. 

The first point has been covered in depth elsewhere \cite{koulakov}, it
will suffice to say that a Delaunay triangulation of a lattice will
produce a unique planar graph. Euler's theorem states that for any
such graph in flat space the following relationship holds between the
number of vertices v, edges e and faces f
\[ 
v+f-e=1.
\]
As a consequence of applying Euler's theorem to a triangular lattice we
can assign a topological charge to each lattice site (or vertex), the
sign and magnitude of the charge depends on by how much the
coordination number (i.e. the number of nearest neighbors the site
has) differs from 6. For example a pentagon has a topological charge
+1 while a square has a charge of +2. Similarly a heptagon has a
topological charge of -1 while an octagon has -2.  Obviously a hexagon
is topologically neutral. The total topological charge for any cluster
is always conserved and must always be equal to +6.

We now show that a change in the density of the lattice leads to a
closure failure of the Burgers circuit. This will allow us to
calculate the Burgers vector density, from which the density of
dislocations and disclinations can be determined.

\begin{figure}
\begin{center}
\includegraphics[width=1.0\columnwidth]{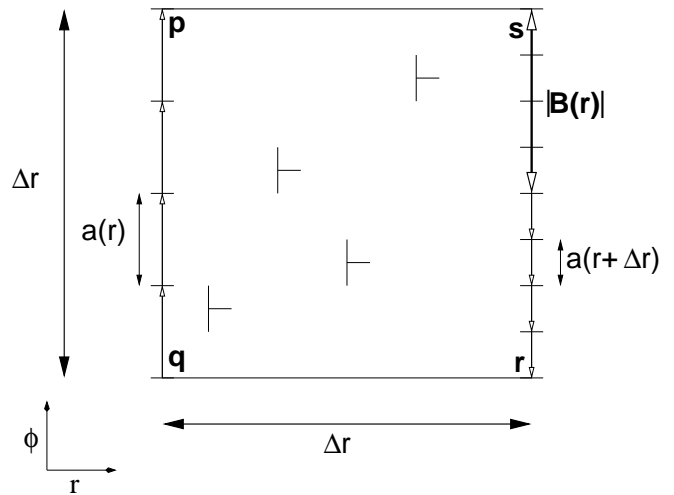}
\caption{Consider a region of crystal in which the lattice density is
  increasing in the r direction. It can be seen that by drawing a
  square Burgers circuit (of dimension $\Delta r$) around the region
  that there are more lattice lines crossing the side {\it sr} than
  crossing the side {\it pq} (while the number of lattice lines
  crossing the side {\it ps} is equal to the number crossing the side
  {\it qr}) The difference in the number of lattice lines leads to a
  closure failure of the Burgers circuit.  This closure failure
  implies that the square contains an excess of dislocations of the
  same sign, the sum of the length of their individual Burgers vector
  is equal to the length of the total Burgers vector. Note that the
  presence of a dislocation is indicated by a T like symbol. The
  vertical bar of the T symbol indicates the side on which the extra
  half plane of atoms is inserted into the lattice, while the
  horizontal top bar indicates the point at which the extra half plane
  of atoms terminates. Thus the Burgers vector is parallel to the top
  bar of the T.}
\label{bcircuit}
\end{center}
\end{figure}

Consider a lattice with smoothly varying density and let us suppose
that the increase in density depends only on the radial distance r.
Drawing a square of dimensions $\Delta r$ around such a region of
lattice, see Fig. \ref{bcircuit}, we see that the number of lattice
rows crossing the side {\it pq} is given by
\[
L(r)=\frac{\Delta r}{a(r)}, 
\]
and the number crossing the side {\it sr} is given by
\[
L(r+\Delta r)=\frac{\Delta r}{a(r+\Delta r)}, 
\]
where a(r), the lattice spacing, is a function of density and in a
triangular lattice is given by
\begin{equation}
a(r)={\sqrt{\frac{2}{\rho(r)\sqrt{3}}}}.
\label{eq:lattice_spacing}
\end{equation}
In constructing a Burgers circuit we take the same number of steps in
the horizontal and vertical directions, however due to the change in
the lattice spacing there will be a closure failure on the side {\it
sr}. Assuming the closure failure is due to an excess of dislocations
of the same sign (which are the origin of the extra half planes),
then the total Burgers vector, due to all the dislocations enclosed,
is given by
\begin{equation}
B_{\phi}(r)=\Delta r -L(r)a(r+\Delta r),
\label{eq:total_burgers_vector}
\end{equation}
where L(r) is the number of steps taken in the Burgers circuit on the
side {\it pq}, (note that the direction of the Burgers vector is
perpendicular to the direction in which the density is increasing).
From Eq. ($\!\!$~\ref{eq:total_burgers_vector}) we have 
\begin{eqnarray}
B_{\phi}(r)
&=&
\Delta r
-L(r)a(r+dr)
\nonumber
\\
&=&
a(r+\Delta r)
\left[
\frac{\Delta r}{a(r+\Delta r)}
-
L(r)
\right]
\nonumber
\\
&=&
\Delta r 
a(r+\Delta r)
\left[
a^{-1}(r+\Delta r) - a^{-1}(r)
\right].
\nonumber
\end{eqnarray}
Taylor expansion to first order yields 
\[
B_{\phi}(r)
=
\Delta r a(r)[\left[\Delta r \frac{da^{-1}(r)}{dr}\right];
\]
to get the Burgers vector density we divide through by the area
of the square and take the limit $\Delta r \rightarrow 0$, giving
\begin{eqnarray}
b_{\phi}(r)
=
\lim_{\Delta r\to\ 0}
\frac{B_{\phi}(r)}{(\Delta r)^2}
&=&
a(r)\frac{da^{-1}(r)}{dr}
\nonumber
\\
&=&
\frac{1}{2}\frac{d}{dr}\ln\rho(r).
\label{eq:bvdensity}
\end{eqnarray}
(Generalizing Eq. ($\!\!$~\ref{eq:bvdensity}), the Burgers
vector density in a 2D plane is given by the vector field \cite{koulakov}
\begin{equation}
{\bf b}({\bf r})
=
a({\bf r})
{\bf \hat{z}}
\times
\nabla
a^{-1}({\bf r}),
\label{eq:general_bvdensity}
\end{equation}
where ${\bf \hat{z}}$ is the unit vector perpendicular to the
surface.) 

The density of the Burgers vector is then 
\begin{equation}
b_{\phi}(r)
=
\pm
\frac{r}{2R^2}
\frac{1}{1-\left(\frac{r}{R}\right)^2},
\label{eq:mybvdensity}
\end{equation}
where $b_{\phi}(r)$ is positive upon substituting Eq.
($\!\!$~\ref{eq:density}) into Eq. ($\!\!$~\ref{eq:bvdensity}) and
negative upon substituting Eq. ($\!\!$~\ref{eq:koulakov_density}) into
Eq. ($\!\!$~\ref{eq:bvdensity}). This then defines a Burgers vector
density field present throughout the lattice. The total Burgers vector
within a radius r can be found by integrating Eq.
($\!\!$~\ref{eq:mybvdensity}) over the area of the disk. 

We have assumed that the changing lattice density is due to the
insertion of extra half planes into the lattice. Thus to get the
density of dislocations, we divide the Burgers vector density by the
distance between crystalline rows, $h=a\sqrt{3}/2$, which gives for
the hard wall case
\begin{equation}
\rho^{H}_d (r)
=
\frac{b_{\phi}(r)}{h}
=
\sqrt{\frac{N}{4\pi \sqrt{3}}}
\frac{r}{R^3}
\frac{1}{(1-\frac{r^2}{R^2})^\frac{5}{4}},
\label{eq:disl_density}
\end{equation}
and there is a similar result for the parabolic case \cite{koulakov}.
Hence, the number of dislocations within a radius r can be found by
integrating Eq. ($\!\!$~\ref{eq:disl_density})  over  the area $r'<r$.

In addition to dislocations we expect disclinations in the lattice.
These are present if the Burgers vector density is rotational. The
density of disclination charge, $\tilde{s}(r)$, is given by the ${\bf \hat{z}}$ 
component of the curl of ${\bf b}({\bf r})$
which reduces in our situation to 
\begin{eqnarray}
\tilde{s}(r)
=
\frac{1}{r}
\frac{\partial}{\partial r}
\left(
rb_{\phi}
\right)
=
\frac{1}{2}
\nabla^{2}
\ln
\rho(r)
=R(r),
\label{eq:curl_of_b}
\end{eqnarray}
where $R(r)$ is the scalar curvature
and is equal to the density of disclination charge. This
quantity is also known in the theory of plasticity as the
incompatibility \cite{kroner}. This relationship can be generalized
to include free disclinations:
\begin{equation}
  \tilde{s}({\bf r})=s({\bf r})-\epsilon_{ik}\nabla_{k}b_{i}({\bf
    r})=R({\bf r}).
\label{eq:disclination_density}
\end{equation}
In analogy to a dielectric the $s({\bf r})$ term is the charge density
of the free disclinations, induced say as a consequence of the topology of the
space in which the lattice is embedded, and
$-\epsilon_{ik}\nabla_{k}b_{i}({\bf r})$ is the polarization
contribution from dislocations \cite{Chaikin}. It turns out that for the 
hard wall case the six disclinations induced by the disk topology
are always at the edge of the system. For the harmonic potential 
they are close to the edge in small systems, but as $N$ increases, the six 
topologically induced disclinations migrate towards the interior. 
Thus  for values of $r$ away from the hard wall, the
  density of
disclination charge, for the hard-wall system, can be found by substituting
 Eq. ($\!\!$~\ref{eq:density}) into
Eq. ($\!\!$~\ref{eq:curl_of_b}), 
\begin{equation}
\tilde{s}({\bf r})
=
\frac{1}{2}{\nabla^2}\ln\rho(r)
=
\frac{1}{R^2}
\frac{1}
{\left(1-\frac{r^2}{R^2}\right)^2}
.
\label{eq:my_disc_density}
\end{equation}

Integrating Eq. ($\!\!$~\ref{eq:my_disc_density}) over the area $r'<r$
 gives the total disclination charge within a radius r
\begin{eqnarray}
\Sigma(r)
&=&
2\pi
\int^{r}_{0}\!\,
r'\tilde{s}(r')
dr'
\\
&=&
\left(
\frac{r}{R}
\right)^2
\frac{\pi}{1-\left( \frac{r}{R} \right)^2}.
\end{eqnarray}
If the lattice only contains disclinations with charge
$\frac{\pi}{3}$, then the number of disclinations within a given radius is
\begin{equation}
N_{disc} (r)
=
\frac{\Sigma(r)}{\pi/3}
=
\left(
\frac{r}{R}
\right)^2
\frac{3}{1-\left( \frac{r}{R} \right)^2},
\label{eq:n_of_disc}
\end{equation}
where for the hard wall confined system we expect the lattice
interior to contain an excess of 7 coordinated disclinations. For 
the harmonically confined system the number of 5
coordinated disclinations induced by the 
changing density can be similarly calculated 
and equals $N_{disc}(r)$.    The fact that both
the density of the Burgers vector Eq. ($\!\!$~\ref{eq:mybvdensity}) and
the disclination charge density due to the changing density 
are equal but opposite for the two systems suggests that they are
``mirror images'' of each other. It makes sense then to compare and
contrast the properties of these two systems.

\subsection{Smooth part of the energy}

In the following we examine the various terms which contribute to the
overall energy of the system of charges in a hard wall confining
potential. A similar study has already been made for the system with
parabolic confinement \cite{koulakov2}. We believe that the energy
will have the form, as $N$ becomes large, of a series in decreasing
powers of $N^{\frac{1}{2}}$:
\begin{equation}
E_{Smooth}=
\kappa_{1}\frac{N^2}{R} 
+
\kappa_{2}\frac{N^{\frac{3}{2}}}{R}
+
\kappa_{3}\frac{N}{R}
+
\kappa_{4}\frac{N^{\frac{1}{2}}}{R}
+
\kappa_{5},
\label{eq:smoothy}
\end{equation}
The first coefficient $\kappa_1=\pi/4$; this is calculated next.  We
have obtained from our numerical estimates of the ground state energy
of systems with varying values of $N$ the following estimates of the
other coefficients: $\kappa_2=-1.562033$, $\kappa_3=0.975852$,
$\kappa_4=-0.008196$ and $\kappa_5=-0.307608$.

The first term is the `electrostatic energy'. It can be calculated by
approximating the density by its continuum limit form, to give
\[
E_{ES}
=
\frac{1}{2}
\int\!\,
\rho(r)
d^2{ r}
\int\!\,
\rho(r')
d^2{ r'}
\frac{1}{|{\bf r} - {\bf r}'|}
;
\]
upon recognizing the second integral as the electrostatic potential
 $\mu$, see 
 Eq. ($\!\!$~\ref{eq:n=1(2d)problem}), this can be written as
\[
E_{ES}
=
\mu
\frac{1}{2}
\int\!\,
\rho(r')
d^2{ r'}
=
\frac{\pi}{4}\frac{N^2}{R}.
\]

The next largest term is the `correlation energy' which is the first
correction to the continuum limit approximation because the charges
are discrete.  Koulakov and Shklovskii \cite{koulakov2} suggested that
it could be estimated by using the local density approximation (LDA),
which states that for a large enough cluster, locally the density can
be assumed to be constant, so the correlation energy of a lattice with
non-uniform density should be the same as the first correction to the
electrostatic energy of an infinite system of uniform density:
\begin{eqnarray}
E_{Corr}
&=&
-\frac{1}{2}\int^{R}_{0}\!\,
2\pi r dr
\rho (r)
\beta
\sqrt{\rho (r)}
\nonumber
\\
&=&
-\beta
\sqrt{\frac{1}{2\pi}}
\frac{N^\frac{3}{2}}{R}
,
\nonumber
\end{eqnarray}
where the value of $\beta$ depends on the geometric properties of the
lattice. For a triangular lattice
$\beta=\beta_{\triangle}=3.921034$ and for a square lattice
$\beta=\beta_{\Box}=3.898598$ \cite{bonsall}. Visual inspection of the
ground states of the system (see section IV.A.5) suggest that locally
the lattice is triangular, thus in calculating the correlation energy
we expect that $\beta=\beta_{\triangle}$.  If this  expectation were correct
it would predict that $\kappa_{2}=-1.5642939$, which is quite close to
our numerical estimate $-1.562033$. We discuss in the following
sections some possible explanations for this small discrepancy and
also how the remaining $\kappa$ coefficients were obtained.

\section{Numerical Results}

In this Section we present the results of our numerical simulations
and compare them with the continuum model developed in Section III.

Simulated annealing experiments were carried out for the system with a
hard wall boundary, as described in Section II, for systems ranging in
size from $N=2$ to $N=100$. Our results either agreed with or gave a
slightly lower energy than those published by Kong et al \cite{kong}.
Simulated annealing experiments were also carried out for larger
systems, $N=150$, 200, 250, 500, 1000, 2000 and 5000. To ensure good
results a range of annealing schedules were tried. Upon finding the
optimal schedule, minimization was repeated as many times as possible,
thus not only improving the chances of finding a good result but also
generating a collection of states which could be used for further
analysis. 

For the system with parabolic confinement we restrict our efforts to
large system i.e. $N=1000$, 2000 and 5000. Numerical experiments have
already been carried out for small systems by Koulakov and
Shklovskii, see \cite{koulakov2} and \cite{koulakov}. Starting with a
random initial configuration of charges, a conjugate gradient
algorithm was used to minimize the energy of the system. In each case
this process was repeated 1000 times and the cluster with the lowest
energy was identified.

For each system the best result was triangulated; this
was done by projecting the charges onto a paraboloid and using the
Delaunay triangulation package Qhull \cite{qhull}. The results were
then displayed using the graphics package Geomview \cite{geomview}. An
additional routine was used to highlight defects in the clusters,
points with five nearest neighbors were colored red, while those
with seven and eight nearest neighbors were colored green and blue
respectively.

\subsection{Hard Wall System}

\subsubsection{Distribution of Charges}

\begin{figure}[t]
\begin{center}
\includegraphics[width=0.8\columnwidth]{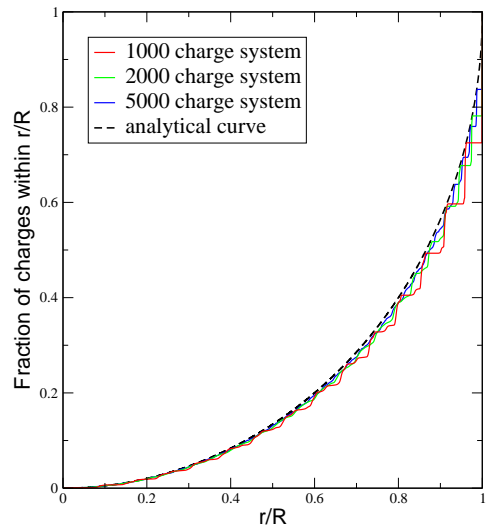}
\caption{Histograms of the fraction of the total charge of the system
  within a given radius $r$. Results for clusters containing 1000, 2000,
  and 5000 charges are colored red, green and blue respectively.  The
  continuum limit result is given by the black dashed line. Note
  that the step-like behavior towards the edge indicates that the
  charge in this region is concentrated into a series of concentric
  shells; the charge arrangements in this area are very different 
   from those in the cluster interior.}
\label{Norm_charge}
\end{center}
\end{figure}

In the continuum limit the charge is distributed according to Eq.
($\!\!$~\ref{eq:Ncharges}); this quantity can be compared with its
actual value in a finite sized cluster which we call $N_{fin}(r)$,
where for a given radius r, $N_{fin}(r)$ is the number of charges
enclosed within that radius.  We choose to compare the integrated
quantity as opposed to the charge density itself as this yields a less
noisy result. Fig. \ref{Norm_charge} gives the fraction of the total
charge enclosed as a function of radius, i.e. $N_{fin}(r)/N$, for the
three largest systems simulated (in each case the result with the
lowest energy is used). Also shown for comparison is the fraction of
charge enclosed in the continuum limit, i.e. $N(r)/N$. By scaling the
charge enclosed in this manner, different sized systems can be easily
compared. The curves in Fig. \ref{Norm_charge} suggest that with
increasing system size the charge distribution approaches the
continuum result.

\begin{figure}[t]
\begin{center}
  \includegraphics[width=0.6\columnwidth,
  angle=270]{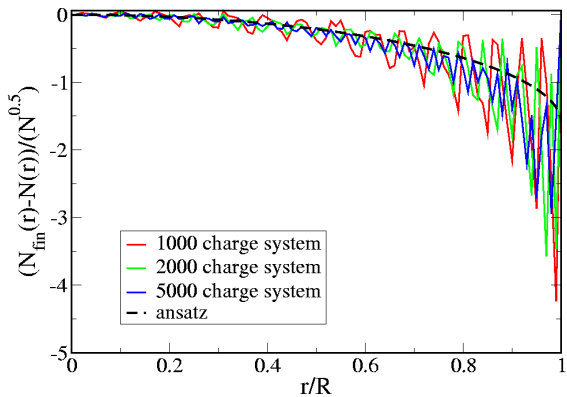}
\caption{ Plot of $\Delta N$ for the three largest systems, scaled by
  $\sqrt{N}$ versus distance $r$ from the center.  $\Delta N$ is the
  difference between the number of charges within a given radius for
  finite systems $N_{fin}(r)$ and the continuum result $N(r)$ (which
  is given by Eq. ($\!\!$~\ref{eq:Ncharges})). The black dotted line is
  an attempt to fit the data by replacing $N$ by $N-CN^{\frac{1}{2}}$
  in Eq. ($\!\!$~\ref{eq:Ncharges}) with $C \approx 1.6$.  }
\label{charge_deviation}
\end{center}
\end{figure}

However, there is a systematic difference between the charge
distribution in the continuum limit and that for finite sized systems,
which is not just a local effect but varies on the scale of the radius
$R$ of the system.  Its presence is revealed on plotting $\Delta
N(r)=N_{fin}(r)-N(r)$, for the three largest systems.  There is a
correction to the continuum expression for $N(r)$, which is of order
$\sqrt{N}$. (The continuum expression for $N(r)$ is of order $N$).
$\Delta N(r)/\sqrt{N}$ is plotted in Fig. \ref{charge_deviation} for
the three largest system sizes which we studied. It shows that there
is a deficiency of charge in the system interior. Furthermore $\Delta
N$ falls to zero only at the edge of the system, which means that the
missing charge is to be found here in the form of extra charges, $N_s$
in number, at the surface. The leading term for the density is the
result in the continuum limit and is of order $N/R^2$ as in Eq.
(\ref{eq:density}); Fig.  (\ref{charge_deviation}) shows that there
are corrections to it of order $N^{\frac{1}{2}}/R^2$. We have been
unable to obtain any analytic understanding of this correction term,
but we have noticed that it can be quite well approximated by
replacing $N$ in Eq.  (\ref{eq:density}) by $N-N_s$ where $N_s
=CN^{\frac{1}{2}}$ and $C \approx 1.6$.  We shall estimate the number
of charges in the outer shell of the system i.e. on the hard wall, and
show that in fact the $\text{``excess''}$ surface charge $N_s$, (which
is of order $N^{\frac{1}{2}}$) is only a small contribution to it at
large $N$ as the number of charges in the outer shell increases as
$N^{\frac{2}{3}}$.

\begin{figure}
\begin{center}
  \includegraphics[ width=0.8\columnwidth ,
  angle=270]{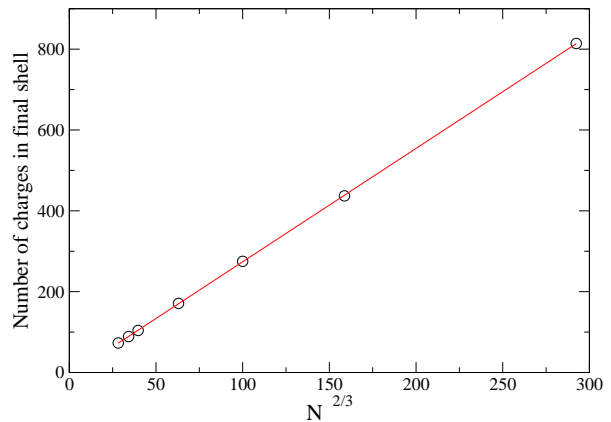}
  \caption{A plot of $N_{fin}(R)$ against
    $N^{2/3}$ for systems in the range N=150 to 5000. To find
    $N_{fin}(R)$ we simply counted the number of charges in the
    outermost shell in each system.}
\label{num_in_final_shell}
\end{center}
\end{figure}

The number of charges we would expect to find
at the edge of the system can be estimated by the following argument:
let the lattice spacing at the edge of the crystal be given by
$a_{e}$, then the number of charges contained within a disk centered on
the origin of radius $R-a_{e}$ is
\[
N(R-a_{e})
=
N
\left(
1
-
\sqrt{
1-
\left(
1-X
\right)^2
}
\right),
\]
where $X={a_e}/{R}$. Expanding the $(1-X)^2$ term to first order
gives
\[
N(R-a_{e})
\approx
N
\left(
1-\sqrt{2X}
\right).
\]
Therefore the number of charges within a distance $a_{e}$ from the edge is
\[
N_{e}=N-N(R-a_{e})
\approx
N\sqrt{\frac{2a_{e}}{R}}.
\]
Imposing the condition that $N_e$ must be equal to the number of
charges on the perimeter $N_p=2 \pi R/ a_e$, yields the ratio
\begin{equation}
\frac{R}{a_e}
\approx
\left(\frac{N}{\pi\sqrt{2}}\right)^{\frac{2}{3}}
\label{eq:a_eff},
\end{equation}
hence the number of charges on the edge, in the continuum limit,
scales with $N^\frac{2}{3}$. To compare with the results of our
numerical simulations we therefore plotted $N_{fin}(R)$, the total
number of charges in the outermost shell of each cluster, against
$N^{2/3}$. As shown in Fig. \ref{num_in_final_shell} there is good
agreement between these estimates and the results of our numerical
simulations.

\subsubsection{The Local Density Approximation}

\begin{figure}
\begin{center}
  \includegraphics[ width=0.7\columnwidth ,
  angle=270]{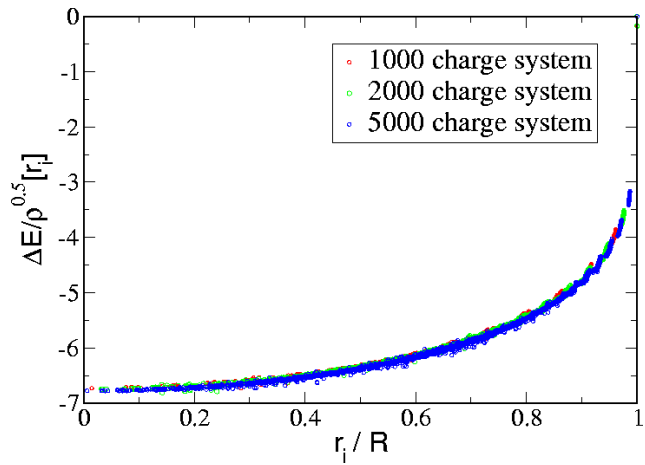}
  \caption{Test of the LDA approximation. If the approximation were perfect $\Delta E(r_i)/\sqrt{\rho(r_i)}$
 would be independent of $r_i/R$ and equal to the numerical value $-3.921034$.}
\label{LDAtest}
\end{center}
\end{figure}

In Section III.C we calculated the correlation energy of the lattice
using the local density approximation (LDA) which is based on 
the idea that locally  the lattice 
appears to be triangular.  Physically this seems to be a very
reasonable approximation, but we observed that the numerical data was
not in perfect agreement with this approximation. In this Section we
shall investigate the matter further.

The correlation energy {\em of a given charge} will be defined as the
difference in energy between the interaction of a charge with all the
other charges in the system after the subtraction of the interaction
of the charge with the continuum approximation for the other charges
\cite{bonsall}:
\begin{eqnarray}
\Delta E(r_i)
&=&
\left[
\sum_{j\neq i}^{N}
\frac{1}{|{\bf r}_i-{\bf r}_j|}
-
\int
\frac{\rho^{H}({r})d^2{r}}{|{\bf r}_i-{\bf r}|}
\right]
\nonumber
\\
&=&
\left[
\sum_{j\neq i}^{N}
\frac{1}{|{\bf r}_i-{\bf r}_j|}
-
\frac{\pi N}{2R} 
\right]
\label{eq:beta_eq_first}
\end{eqnarray}
In the LDA, this quantity would be expected to equal
$-\beta_{\triangle} \sqrt{\rho(r_i)}$. In Fig. (\ref{LDAtest}) we have
plotted $\Delta E(r_i)/\sqrt{\rho(r_i)}$ against $r_i/R$ to test this
expectation. The agreement is non-existent.

The origin of this discrepancy is not hard to find. It arises because
$\Delta E(r_i)$ is a quantity of size $N^{\frac{1}{2}}/R$.
$\sqrt{\rho(r_i)}$ is of this magnitude, but so are also the
contribution of the deviations of the density from $\rho^H(r)$ which
are of order $N^{\frac{1}{2}}/R^2$. Alas, we do not have a calculation
of these deviations. However, we have found that our attempts to model
them by changing $N$ in the expression for $\rho^H(r)$ to
$N-C\sqrt(N)$ and allowing for the compensating surface charge needed
for charge neutrality does at least reduce the discrepancy in Fig.
(\ref{LDAtest}). So it could be that with proper allowance for these
corrections to the density, the LDA might still be valid.

\subsubsection{Distribution of Disclinations}

\begin{figure}[]
\begin{tabular}{cc}
\includegraphics[width=1.0\columnwidth]{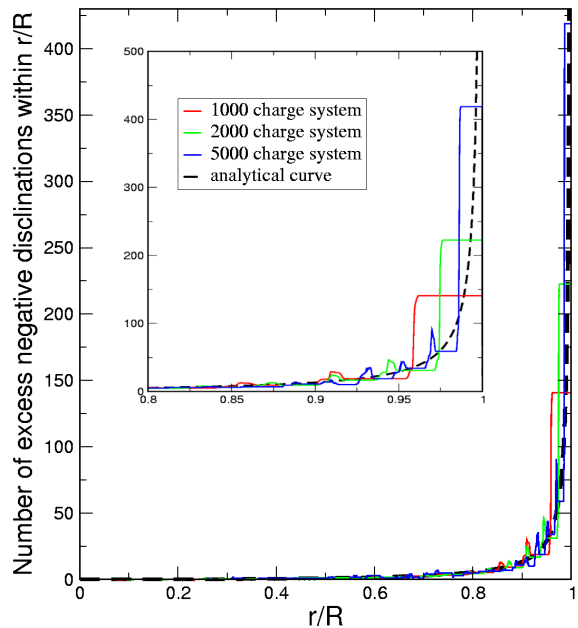}
\end{tabular}
\caption{Histogram giving the total number of disclinations within a given
radius. Clusters containing 1000, 2000, and 5000 charges are colored
red, green and blue respectively. The analytical curve is given by the
black dashed line. The inset is a magnification giving the last 20\%
of the graph showing there is good agreement right up to the edge.}
\label{disclination_histogram}
\end{figure}

The number of excess disclinations located within a given
radius is predicted  by Eq. ($\!\!$~\ref{eq:n_of_disc}). To compare this
 with our simulations we use the following method: for a given
radius, we count the number of positive and negative disclinations
enclosed, where we expect from Eq. ($\!\!$~\ref{eq:n_of_disc}) that in
the interior of the lattice the number of negative disclinations will
always be greater than the number of positive disclinations. We adopt
the convention that a 7 coordinated point counts as one negative
disclination, an 8 coordinated point as two negative disclinations and
a 5 coordinated point counts as one positive disclination.  In
Fig. \ref{disclination_histogram} we plot the number of excess negative
disclinations within a given radius and compare with
Eq. ($\!\!$~\ref{eq:n_of_disc}). It is evident that there is
convergence with increasing system size. For every negative
disclination in the lattice interior there is a compensating positive
disclination on the lattice edge. In addition, the lattice edge also
contains 6 extra positive disclinations due to Euler's theorem.

\subsubsection{Cluster Energies}

\begin{figure}
\begin{center}
\includegraphics[width=0.9\columnwidth]{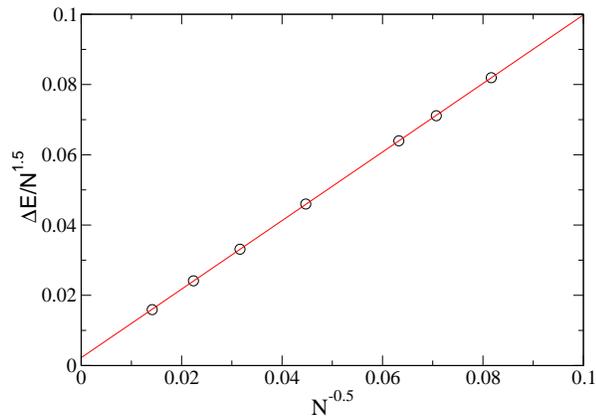}
\caption{$\frac{\Delta E}{N^\frac{3}{2}}$ against $N^-\frac{1}{2}$ plotted for
N=150 to N=5000. The red line is a fit
of the form y=0.975852x + 0.00223295}
\label{diff_cont_num_E_fig}
\end{center}
\end{figure}

In this section we compare the scaling of the energy as given in section
III.C with the energies found by simulated annealing. 

Having identified the electrostatic and correlation energies as the
first two leading order contributions to the energy of the cluster,
which scale as $N^2$ and $N^{1.5}$ respectively, we make the ansatz
that the remaining terms in the series descend in powers of $\sqrt{N}$,
thus
\begin{eqnarray}
E_{Smooth}
&=&
\frac{\pi}{4}\frac{N^2}{R}
+
\left
(
\alpha
-
\beta_{\triangle}
\sqrt{\frac{1}{2\pi}}
\right
)
\frac{N^\frac{3}{2}}{R}
\nonumber
\\
&+&
\kappa_3\frac{N}{R}
+
\kappa_4 \frac{N^{\frac{1}{2}}}{R}
+
\kappa_5.
\label{eq:E_Smooth}
\end{eqnarray}
The second term is divided into two parts:
$\beta_{\triangle}$ is the correlation energy if the lattice is
strictly triangular everywhere and the LDA applies; the term containing the parameter
$\alpha$ is a correction to this assumption.

We now compare the scaling of the energy given by
Eq. ($\!\!$~\ref{eq:E_Smooth}) with the energies found by simulated annealing.
Since the electrostatic and correlation/surface terms dominate for
large systems, it is natural to ask what influence the remaining terms
may have, so we plot
\begin{equation}
\frac{\Delta E}{N^\frac{3}{2}} 
=
\frac{1}{N^\frac{3}{2}}
\left(
E
-
\frac{\pi}{4}\frac{N^2}{R}
+
\beta_{\triangle}
\sqrt{\frac{1}{2\pi}}
\frac{N^\frac{3}{2}}{R}
\right)
\label{eq:diff_cont_num_E_eq}
\end{equation}
against $N^{- \frac{1}{2}}$, see Fig.  \ref{diff_cont_num_E_fig}.  From
the intercept we find $\alpha=0.00223295$ and from the gradient
$\kappa_3=0.975852$. Let us replace the coefficients multiplying the
$N^{1.5}/R$ term in Eq. ($\!\!$~\ref{eq:E_Smooth}) with
\begin{equation}
\kappa_2=\sqrt{\frac{1}{2\pi}}\beta_o=\alpha-\sqrt{\frac{1}{2\pi}}\beta_{\triangle},
\label{eq:beta_0}
\end{equation}
where $\beta_o = 3.915436 $ is the `observed' value of $\beta$.
 It differs from $\beta_{\triangle}$  by
approximately 0.15\%. This  might be due
to a failure of the LDA approximation, but it might also be due to the
fact that we are almost certainly not obtaining the ground states for
each value of $N$, which has uncertain consequences for the accurate
determination of the coefficient $\kappa_2$.

\begin{figure}
\begin{center}
\includegraphics[width=0.8\columnwidth, angle=270]{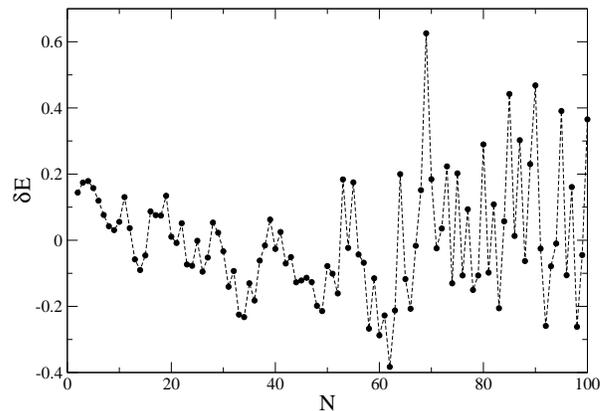}
\caption{A plot showing the difference between the energy for small
  clusters found by simulations and the smooth part of the energy.}
\label{fluct}
\end{center}
\end{figure}

There is no appreciable bending of the curve shown in Fig. 
(\ref{diff_cont_num_E_fig}), which suggests that the remaining terms in
Eq. ($\!\!$~\ref{eq:E_Smooth}) are insignificant in the large N limit.
(The coefficients $\kappa_3$, $\kappa_4$ and $\kappa_5$ were actually
determined by using a curve fitting package in the range $N=2$ to
$100$).  Furthermore as all the data points lie almost perfectly on a
straight line, we conclude that the numerical results are fairly
reliable. In fact for large systems, this plot often enabled us to
determine which clusters generated by the simulated annealing
algorithm required further annealing. These were the clusters for
which the data points were slightly above the fit plotted in Fig.
\ref{diff_cont_num_E_fig}.

\begin{figure*}[htp]
  \begin{center}
    $\begin{array}{c@{\hspace{0.5cm}}c}
      \epsfxsize=8cm
      \epsffile{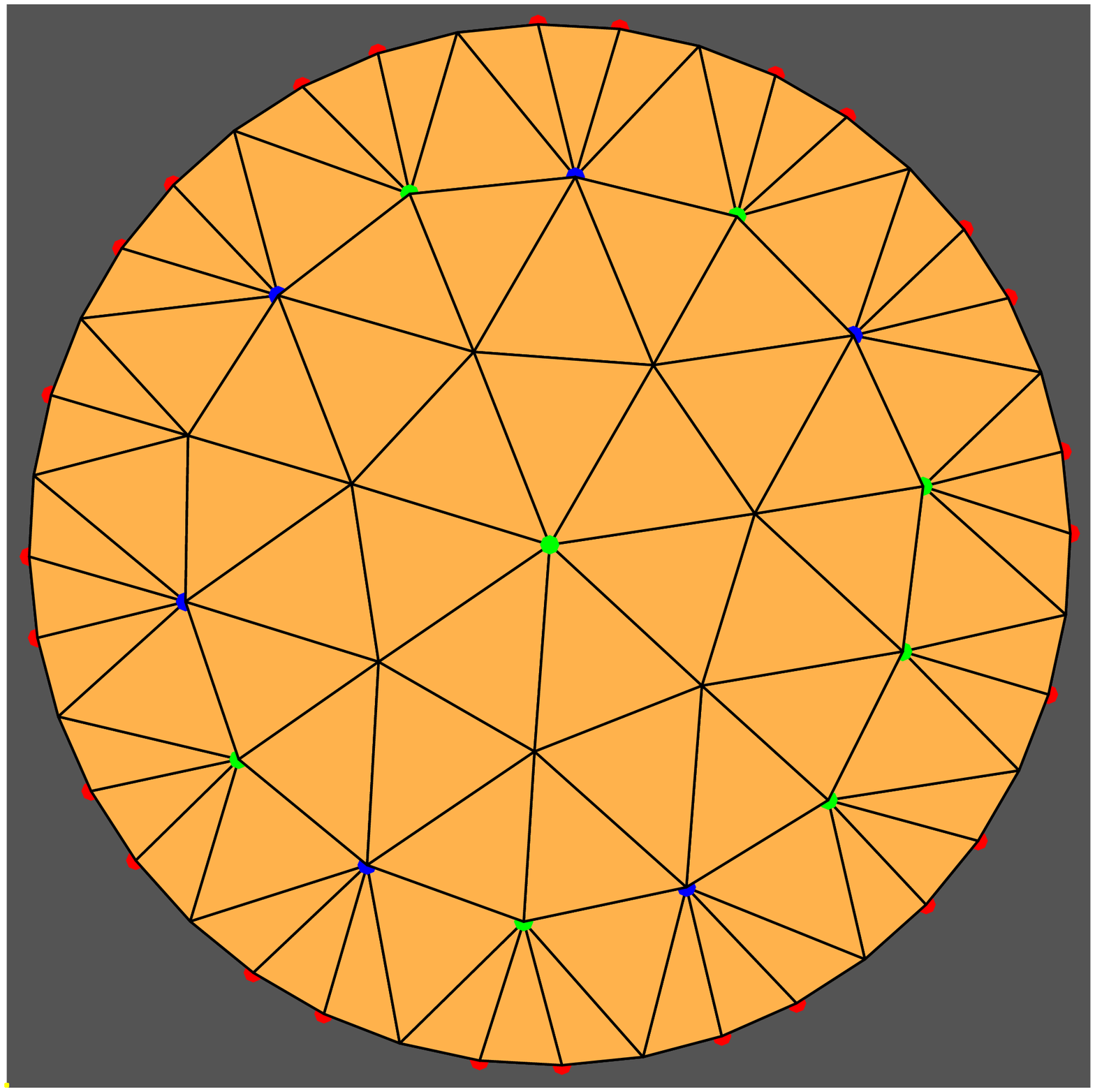} &
      \epsfxsize=8cm
      \epsffile{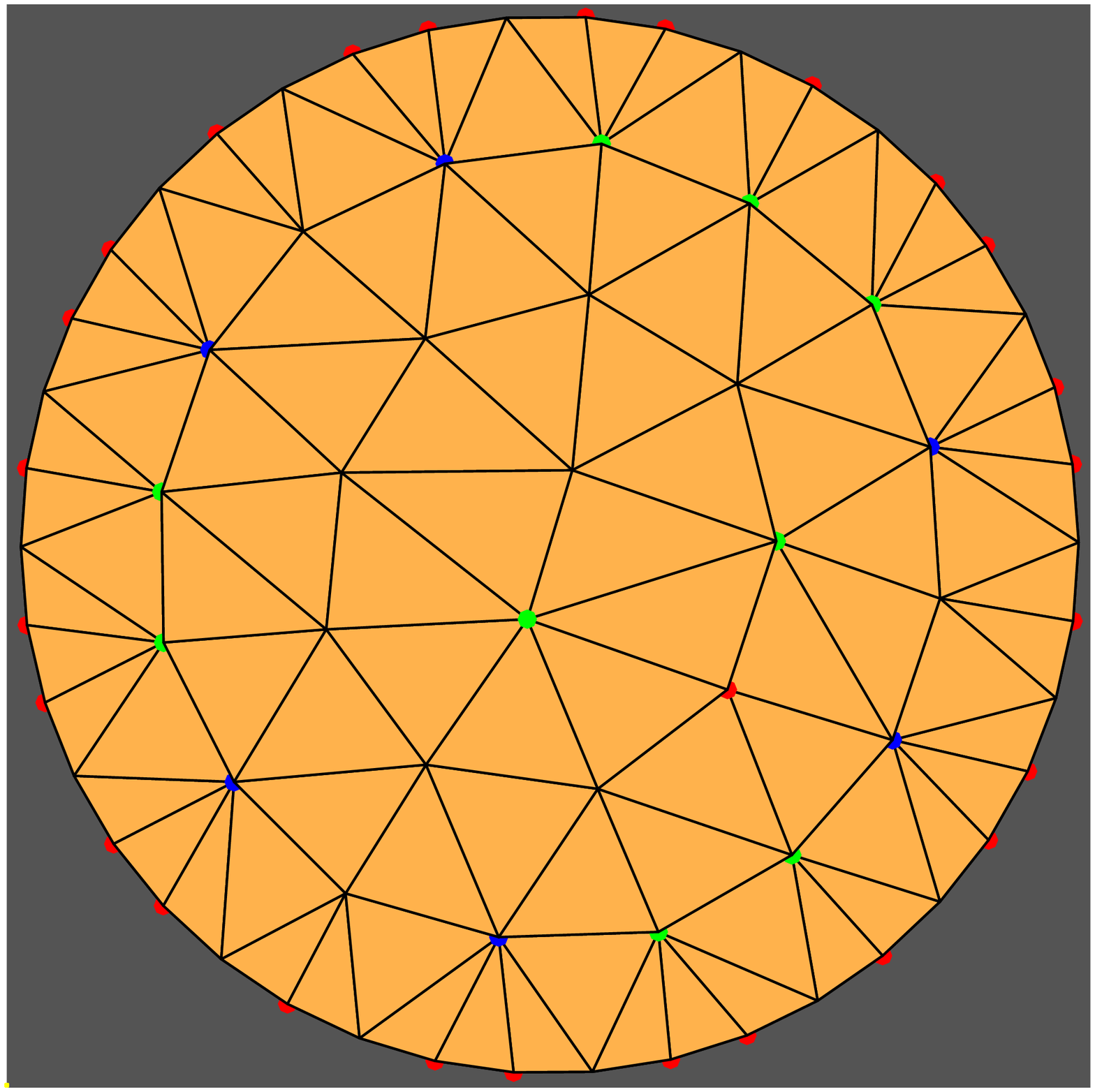} \\ [0.3cm]
      \mbox{\bf 62 Charges} & \mbox{\bf 69 Charges} \\ [0.3cm]
      \epsfxsize=8cm
      \epsffile{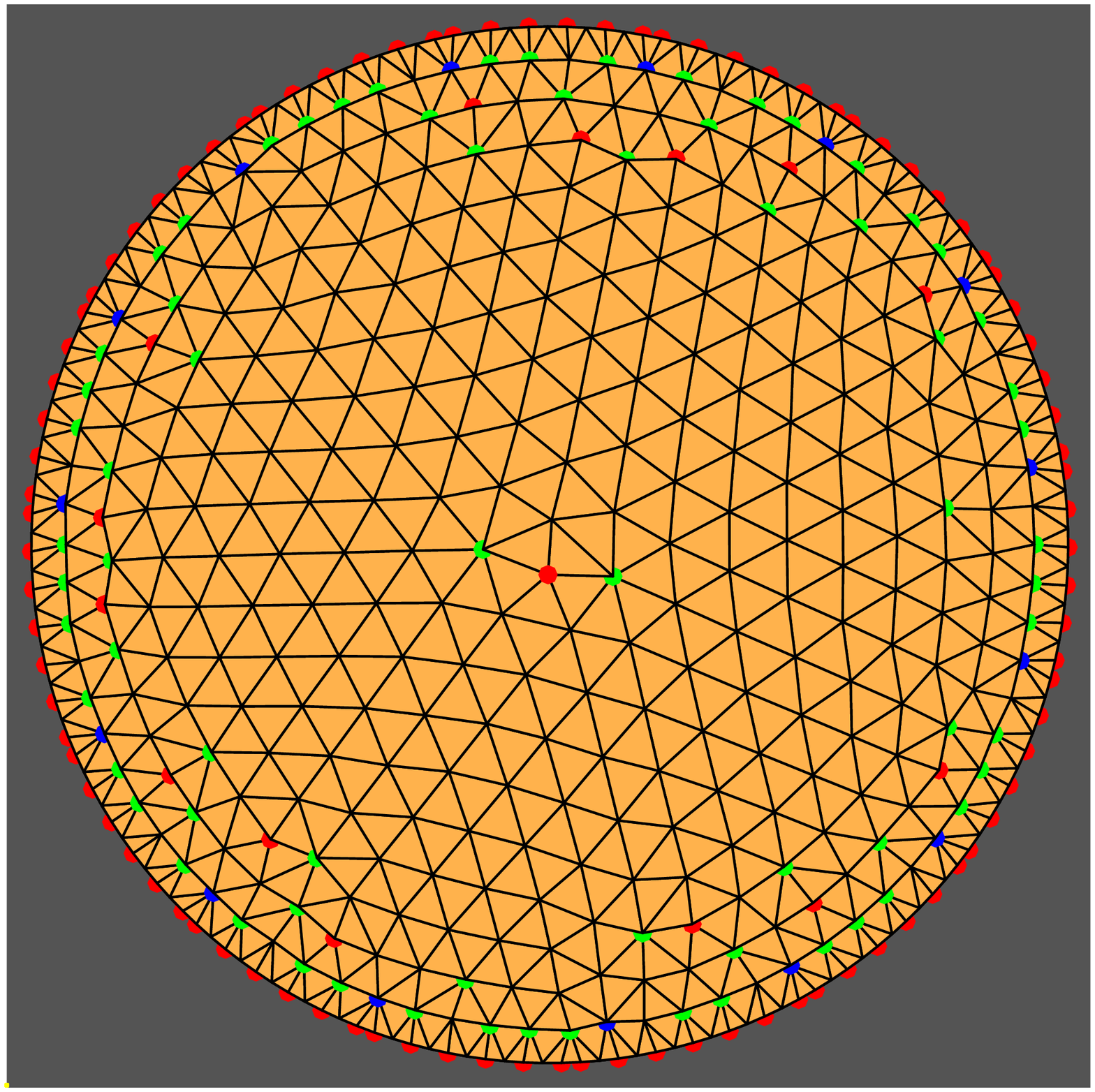} &
      \epsfxsize=8cm
      \epsffile{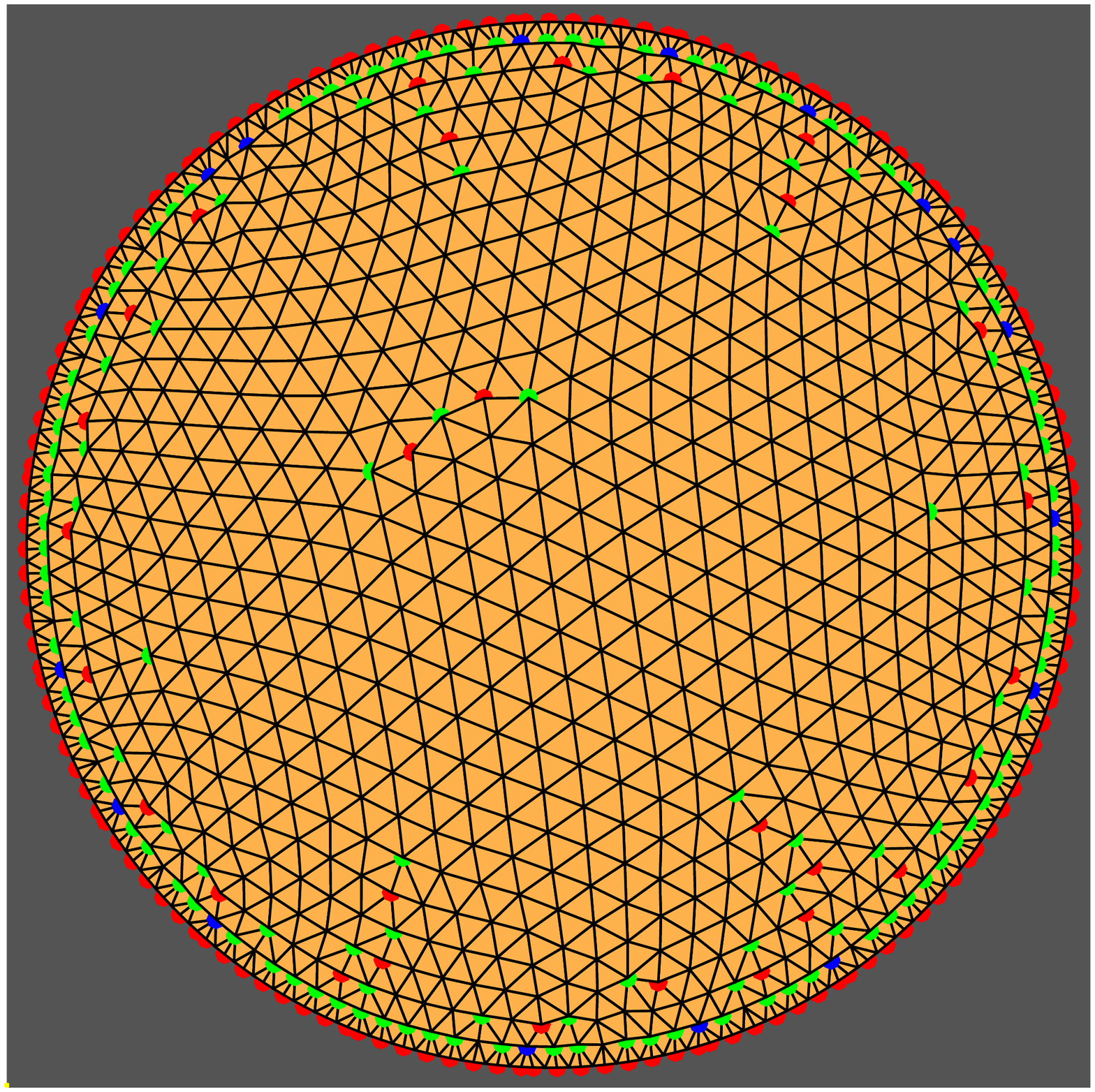}\\ [0.3cm]
      \mbox{\bf 500 Charges} & \mbox{\bf 1000 Charges}
    \end{array}$
  \end{center}
  \caption{{\bf Clusters with a hard wall confining potential.} The
    cluster with $N=62$ has the lowest measured value of $\delta E$,
    while the cluster with $N=69$ has a particularly high value. This
    is because the $N=62$ cluster possesses a high degree of circular
    symmetry, which is commensurate with the hard wall boundary. The
    lack of circular symmetry in the $N=69$ cluster is evident from
    the off-center negative disclination. In addition the boundary
    imposes a shell-like order towards the edge of the cluster, which
    is indicated by the presence of a large number of seven and eight
    coordinated disclinations. For larger systems the shell-like order
    is confined to within a few lattice layers of the edge, the
    lattice interior is composed of a triangular lattice with some
    charged grain boundaries.}
\label{small_systems}
\end{figure*}

Having determined the form of the ``smooth'' part of the energy we can
determine the fluctuating part of the energy, which is defined as
$\delta E = E-E_{Smooth}$. The results are shown in Fig. \ref{fluct}.
Some clusters have a particularly low value of $\delta E$ such as
$N=34$, 49 and 62. These clusters are known as ``magic number states''
and posses a high degree of geometrical symmetry. To illustrate this
point Fig. (\ref{small_systems}) shows a Delaunay triangulation of the
system with 62 and 69 charges, which correspond to the lowest and
highest points, respectively, in Fig.  (\ref{fluct}).

\subsubsection{Small, Medium and Large Clusters}

\input{epsf}

\begin{figure*}[htp]
\begin{center}
\includegraphics[width=12cm, height=12cm]{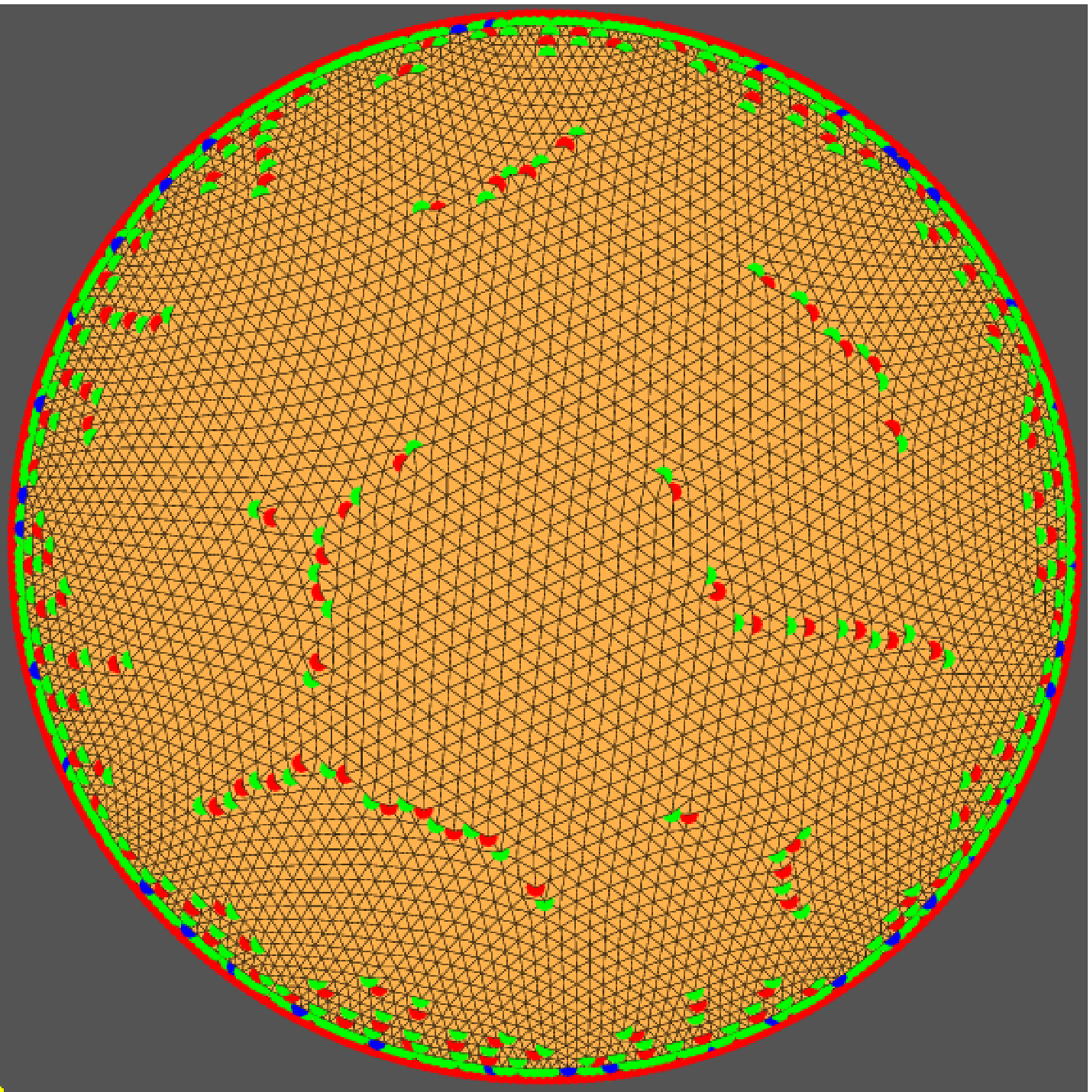}
\\ [0.3cm]
\mbox{\bf 5000 Charges}
\caption{{\bf Clusters with a hard wall confining potential.} The
  remarkable feature of this system is the arched like structure
  towards the edge. This curvature is due to an excess of negative
  disclinations in the lattice interior.  Most of the charged grain
  boundaries appear to be located in the region between the relatively
  undistorted triangular lattice and the arches. We believe that the
  grain boundaries act to screen off the two incommensurate regions.}
\label{very_large_systems}
\end{center}
\end{figure*}

Small clusters, those which contain less than 100 particles, are
dominated by the circular hard wall boundary. In particular, clusters
containing less than 50 charges show no hint of hexagonal order.
Instead the charges are arranged in concentric shells, where many of
the charges have 7 or 8 nearest neighbors. We identify this type of
ordering as ``shell-like''; it is present at the edge of all the
systems. For systems containing more than 50 particles, a region of
the cluster free from the influence of the hard wall boundary begins
to emerge; this region looks like a triangular lattice. At this point
there is a competition between shell-like order and the emerging
hexagonal order. The clusters with the lowest energy are the ones
which can satisfy both simultaneously.  Both $N=62$ and $N=69$ are
clusters which contain a region where most of the charges have six
nearest neighbors. The charges near the center in both clusters can be
considered to be part of a disclination as the central charge has a
different coordination number compared to its neighbors.  Of all the
clusters, $N=62$ has the lowest value of $\delta E$ while $N=69$ has
the highest. The difference is that for N=69 the disclination is
off-center which breaks circular symmetry while in the case of $N=62$,
the disclination is perfectly in the center of the cluster and is
surrounded by a ring of charges which have six nearest neighbors.

For the medium sized clusters, i.e. $100 \leq N \leq 1000$, the influence
of the hard wall boundary on the cluster is contained within a narrow
annular region at the edge. This again is the shell-like region and
contains a large number of seven and eight coordinated disclinations,
which have the effect of destroying the hexagonal order. On the other
hand, the internal region of the lattice is comprised of a relatively
undistorted triangular lattice lattice. The lattice interior contains an excess
of negative disclinations.  However, for every excess negative
disclination in the interior there is a compensating positive one on
the lattice edge. Using the rule that a charge with 7 or 8 nearest
neighbors is a disclination with topological charge -1 and -2
respectively, while a charge with 5 nearest neighbors has charge +1,
the total topological charge of all the clusters was found to be +6.
The 6 positive disclinations due to Euler's theorem are also located
on the lattice edge, hence Euler's theorem is satisfied.

Also with increasing system size isolated disclinations become less
common and are replaced by small topologically charged grain
boundaries; these are chains of alternating positive and negative
disclinations, which contain in total one excess negative
disclination. As shown in Fig. (\ref{disc_close_up}), the alternating
Burgers vectors of the charged grain boundary cancel out and the
overall arrangement constitutes a disclination.

Large clusters are those which contain more than 1000 charges, see
Fig. (\ref{very_large_systems}). In this regime, the interior contains
large areas of lattice separated by charged grain boundaries, which
are  numerous and long. As will be shown in Section V.A, excess
disclinations of one sign in the lattice interior generate lattice
curvature, the effect of which is to bend the lattice lines into a
series of arches. This can be seen with great clarity in the $N=5000$
system. For this system, the bending is sufficiently strong to make
the arched structure incompatible with the ordinary triangular lattice.
It would seem that the grain boundaries arrange themselves to screen
off the two regions.

\subsection{Harmonically Confined System}

\begin{figure}
\begin{center}
\includegraphics[angle=270, width=1.0\columnwidth]{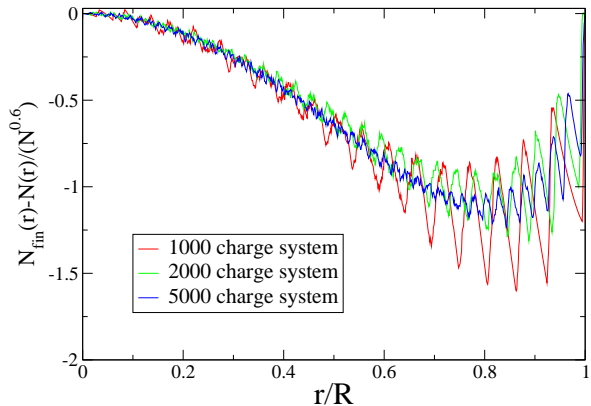}
\caption{Plot of $\Delta N$ for the three largest systems with a
  harmonic confining potential, scaled by $N^{0.6}$ versus distance
  $r$ from the center.}
\label{KScorrection}
\end{center}
\end{figure}

In this Section we consider a 2D cluster of charges confined by a
harmonic confining potential. This system has been studied in
considerable detail by Koulakov and Shklovskii, see \cite{koulakov}
and \cite{koulakov2}, and here we expand on their work.

As for the system with the hard wall confining potential, the actual
density differs from the continuum limit result by subdominant
contributions which are difficult to quantify. Plotting $\Delta
N=N_{fin}(r)-N^P(r)$ for the three largest systems shows that there is
a deficiency of charge in the system interior which is compensated for
by a shell of excess charge at the edge of the system.  A good
collapse of the data for different values of $N$ was found as in when
we plotted
\begin{equation}
\frac{\Delta N}{N^{0.6}}=\frac{N_{fin}(r)-N^P(r)}{N^{0.6}},
\label{eq:kkkp}
\end{equation}
but the origin of this correction with the power $0.6$ is a mystery to us. 

\begin{figure}
\begin{center}
\includegraphics[width=0.8\columnwidth , angle=270]{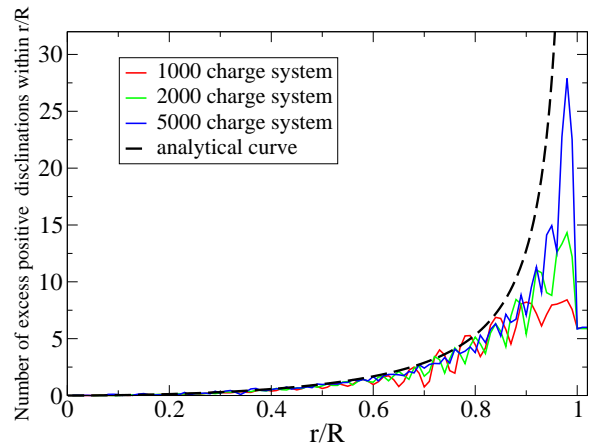}
\caption{ A plot of the number of excess positive disclinations within
  a distance r/R  where we use the rule that a five coordinated and a
  seven coordinated disclination count as a single positive or
  negative disclinations respectively. Results for clusters containing
  1000, 2000 and 5000 clusters are colored red, green and blue
  respectively. The analytical curve is given by the black dashed
  line. Note that at the edge of the cluster the number of excess
  positive disclinations falls to +6. Unlike the number of excess
  disclination in the hard wall system, the number in this system lags
  behind the continuum value significantly.}
\label{koulakov_disc_histograms}
\end{center}
\end{figure}

In the continuum limit, the density of excess disclinations
and the density of the Burgers vector for this and the hard wall
systems are the same except for the sign. It follows that the number
of excess disclinations within a given radius is the same for the two
systems.  A comparison with our numerical simulations is given in
Fig. (\ref{koulakov_disc_histograms}).

\input{epsf}
\begin{figure*}[htp]
  \begin{center}
    $\begin{array}{c@{\hspace{0.5cm}}c}
      \epsfxsize=8cm
      \epsffile{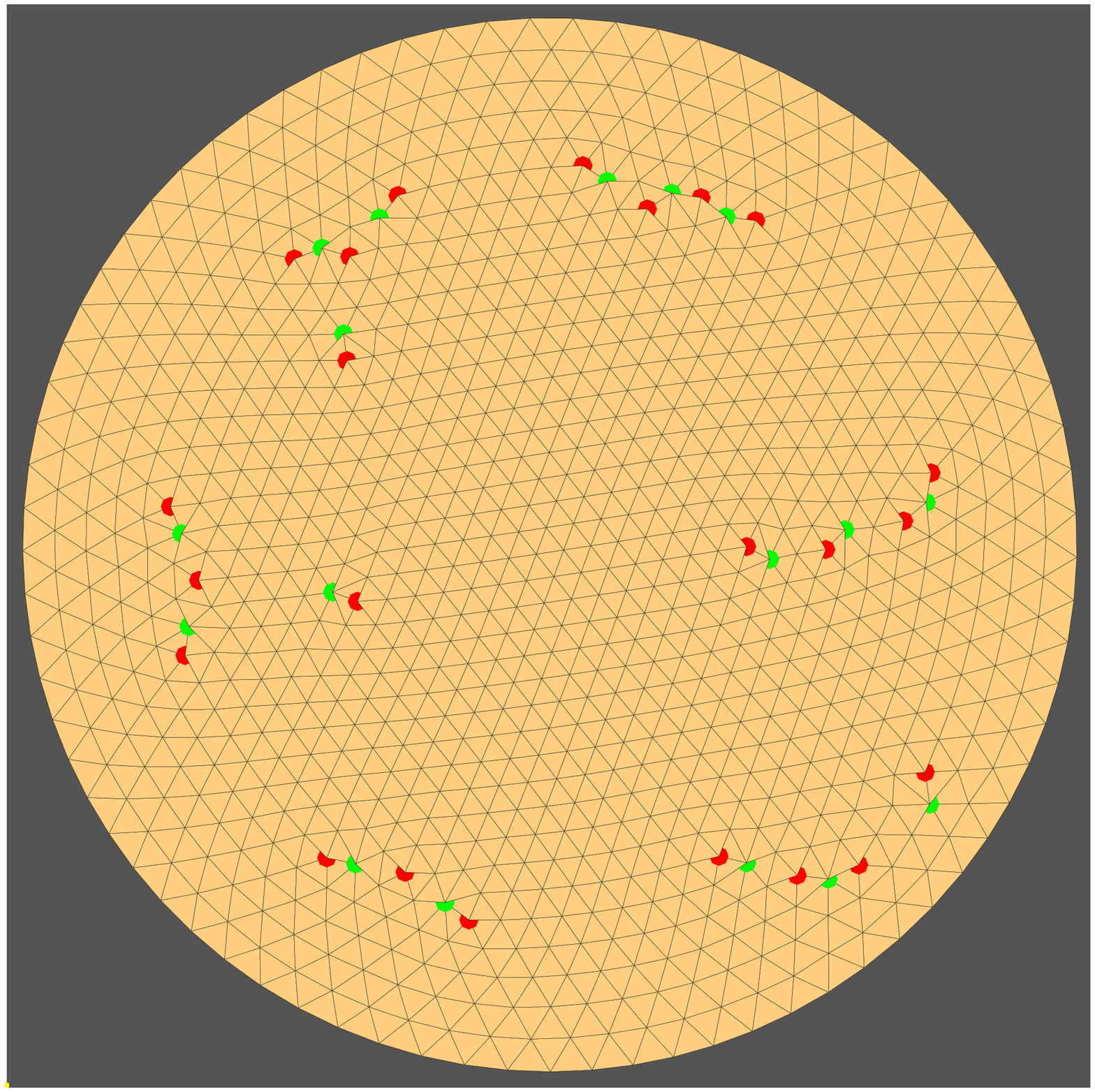} &
      \epsfxsize=8cm
      \epsffile{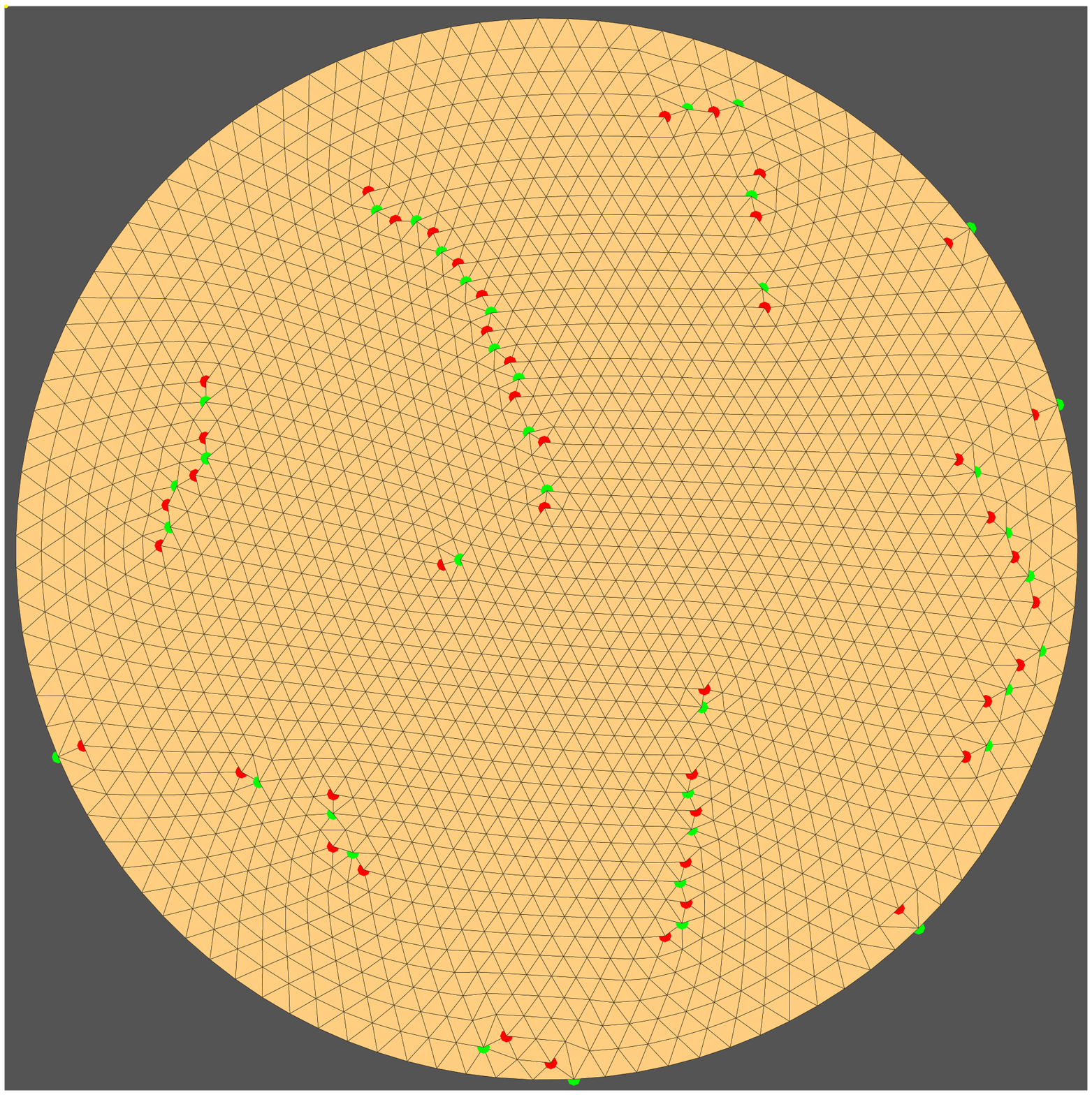} \\ [0.3cm]
      \mbox{\bf 1000 Charges} & \mbox{\bf 2000 Charges} \\ [0.3cm]
    \end{array}$
  \end{center}
  \begin{center}
    \includegraphics[width=12cm, height=12cm]{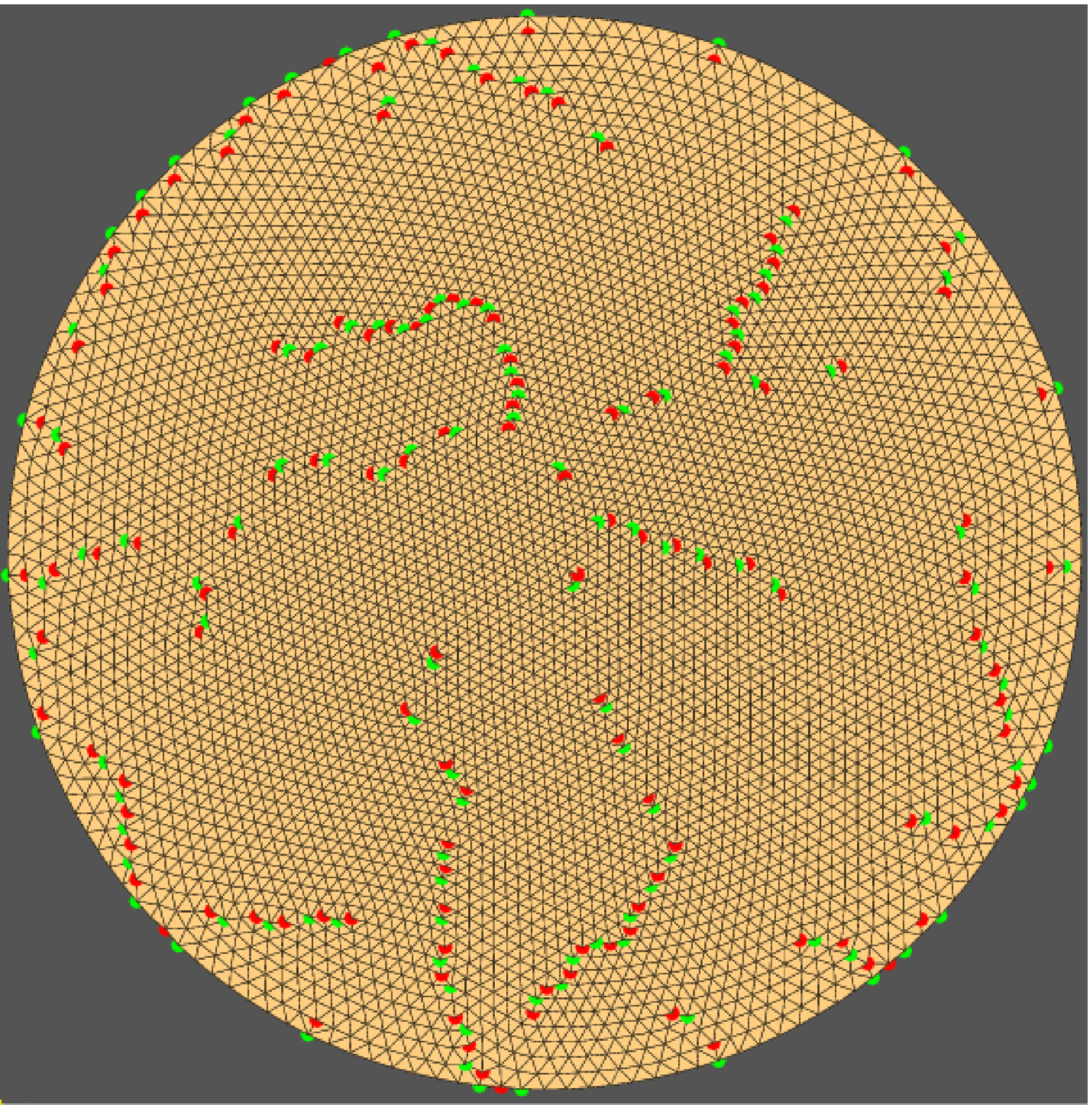}
    \\ [0.3cm]
    \mbox{\bf 5000 Charges}
  \end{center}
  \caption{{\bf Clusters with a harmonic confining potential.} {\it 1000
      charges}: This cluster contains both charged grain boundaries
    and also some isolated dislocations.  These dislocations are not
    involved in screening, but serve to relax the strain energy.  {\it
      2000 charges}: With increasing system size, the charged grain
    boundaries continue to grow in length. Also the isolated
    dislocations become more numerous, especially near the edge of the
    system. Note these dislocations are oriented with the 5
    coordinated disclination pointing towards the center of the
    system. There is a rapid rise in the number of positive
    disclinations just before the lattice edge which are canceled
    by the negative disclinations close to the edge. {\it 5000
      charges}: For very large systems, the additional dislocations
    required to accommodate the changing density
     form uncharged grain boundaries (i.e grain
    boundaries with no over-all topological charge). Clear examples of
    such grain boundaries can be seen towards the edges of the
    cluster. }
  \label{koulakov_systems}
\end{figure*}

Koulakov and Shklovskii demonstrated that provided the cluster is
small $(N \leq 150)$ then there exist certain magic number states
which contain only 6 five-coordinated disclinations demanded by
Euler's theorem. These
disclinations are not right on the lattice edge but are close to it.
 Beyond this limit $(N \geq 150)$ the disclinations are
always accompanied by a screening cloud of dislocations. It was
further shown that provided the cluster is not too large $(N \leq
700)$ then the only defects in the lattice are the disclinations and
their screening clouds, which together form 6 small separate
topologically charged grain boundaries. In fact provide that $N \leq
700 $, the lattice can be divide into an inner region and an outer
region, where the boundary between the two is given by the radius at
which the charged grain boundaries are located. The inner region is a largely 
undeformed triangular lattice while in the outer region the lattice
lines are curved.

Beyond this limit ($N \geq 700$) dislocations not associated with
screening begin to appear, these dislocations are present in order to
ensure that Eq. ($\!\!$~\ref{eq:my_disc_density}) is satisfied,
consequentially the lattice cannot be divided so neatly into two
separate regions \cite{koulakov}.

The numerical work of Koulakov and Shklovskii only dealt with systems
containing less than 700 charges. Our work is concerned with the
behavior of the system as the cluster grows beyond this limit.  An
examination of the cluster containing 1000 charges, see
Fig. (\ref{koulakov_systems}), shows that in addition to the topologically
charged grain boundaries the system also contains a few isolated
dislocations. These dislocations are orientated so that the 5
coordinated disclination points towards the center of the lattice
while the 7 coordinated disclination points radially outwards.  With
increasing system size the general trend is that the screening cloud
around the disclinations continue to grow in length. In addition the
isolated dislocations become increasingly numerous, see the cluster
containing 2000 charges in Fig. (\ref{koulakov_systems}).  Eventually in
addition to isolated dislocations we can observe uncharged grain
boundaries (i.e extended chains consisting of alternating
disclinations but which have no overall topological charge), see the
system containing 5000 charges, Fig. (\ref{koulakov_systems}).

Even though the continuum disclination charge density is the same for
the hard wall system and the system with harmonic confinements,
nevertheless there is a remarkable difference between the two in their
approach to the continuum limit with increasing system size. Compared
to the hard wall system, the number of disclinations in this system is
far fewer. From Fig. (\ref{koulakov_disc_histograms}) we can see that the
number of disclination within a given radius lags behind the continuum
value, while the opposite is true in the hard wall system. For
instance, for 5000 charges in a harmonic trap the number of excess
disclinations at the edge is about 25, while in the hard wall case
this is 400. Even though the curvature of lattice lines in the
continuum limit (which is given by the density of Burgers vector) is
the same for the two systems, the curvature of the lattice lines
appears to be far less pronounced for the system with harmonic
confinement. This may be due to the lack of disclinations, since the
curvature depends on the number of disclinations enclosed within a
radius, see Eq. ($\!\!$~\ref{eq:kurvature}).

\begin{figure}
\begin{center}
\includegraphics[width=0.6\columnwidth]{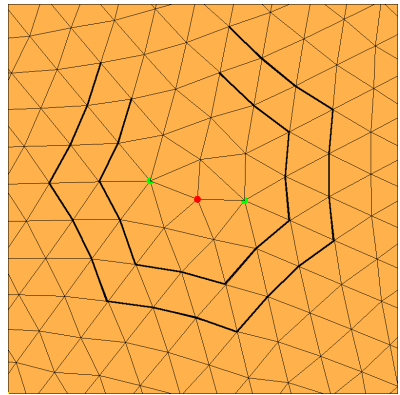}
\caption{ A close up of a small grain boundary with a total
  topological charge of -1. The arrangement can also be thought of as
  a pair of dislocations with opposite Burgers vectors which cancel.
  Drawing a Burgers circuit around the grain boundary leads to a
  closure failure which, like a disclination of charge -1, increases
  with distance from the center.}
\label{disc_close_up}
\end{center}
\end{figure}

Unlike the hard wall system, which requires an excess of 7 coordinated
disclinations in the interior, the harmonic system already contains
six free 5 coordinated disclination due to Euler's theorem. In the
case of the hard wall system these topologically induced 
disclinations are pushed to the
edge of the system while in the harmonic case these disclinations sink
into interior and help in accommodating the non-uniform density. In
contrast to the hard wall system, the density of free disclinations,
given by $s(r)$ in Eq. ($\!\!$~\ref{eq:disclination_density}), cannot
be ignored. From Fig. (\ref{koulakov_disc_histograms}) it can be
seen that up to $r=0.8R$ the total disclination charge does
not exceed +6. Thus up to this point the mechanism by which the
lattice adapts to the decreasing density towards the edge of the
system depends on the arrangement of these free disclinations. In
order for the total disclination charge to match that given by Eq.
($\!\!$~\ref{eq:my_disc_density}), dislocations arise to screen the
disclinations. These screening dislocations have the effect of
smearing out the disclination charge.  To explain what we mean by
this, consider the following qualitative argument. To define a
disclination we must be able to draw a Burgers circuit around it. The
smallest Burgers circuit around a single disclination has a radius
equal to the local lattice spacing. This then defines the minimum size
of the defect. By screening the disclination to form a charged grain
boundary, of the type shown in Fig. (\ref{disc_close_up}), the radius
of the Burgers circuit needed to enclose the object becomes larger,
hence the size of the defect is increased. However, the total
disclination charge enclosed is still the same. Thus the total density
of disclination charge, i.e.  $\tilde{s}({\bf r})$, is reduced. Or to
put it another way, unlike an isolated disclination, which is a single
point in the lattice, the charged grain boundary is an extended
object. However, the total charge of the grain boundary is still $\pm
1$. Thus we can consider this charge to be smeared out over its
length.

Beyond $r=0.8R$  there is a sharp increase in
the number of dislocations, which for the largest systems condense
into uncharged grain boundaries. These dislocations are oriented such
that the 5 coordinated disclination points inwards while the 7
coordinated disclination is on the lattice edge. Thus towards the edge
of the lattice there is a sudden jump in the number of excess positive
disclinations followed by an equally sudden fall. Perhaps as the
number of charges in the system is increased further, these
dislocations might unbind so that  the 7-coordinated disclinations are
pushed to the lattice edge while the 5-coordinated disclinations
remain in the interior.

\section{discussion}

\subsection{Lattice Curvature}

\begin{figure*}
\begin{center}
\includegraphics[width=2.0\columnwidth]{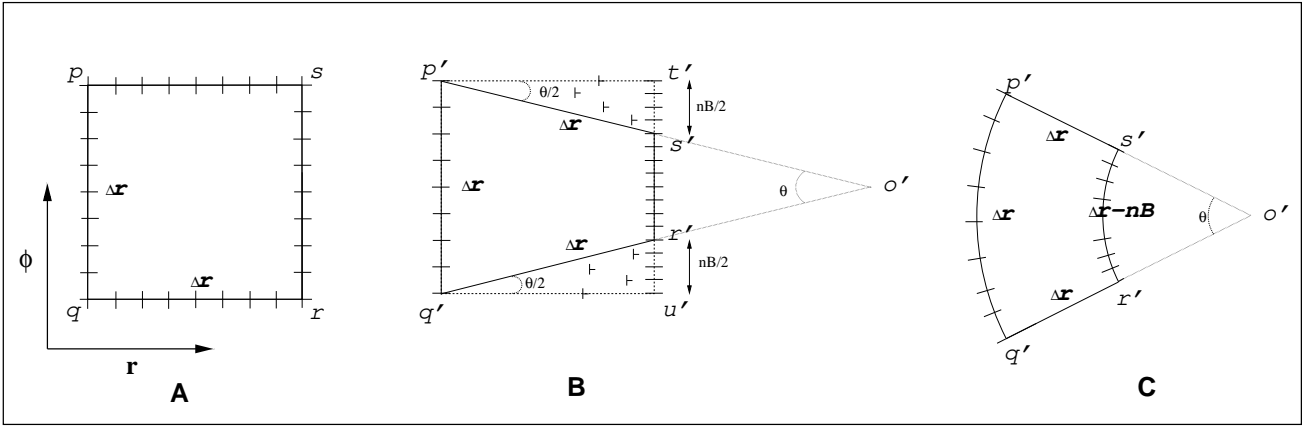}
\caption{ {\it [A]:} An undeformed lattice of dimensions $\Delta r$.
  {\it [B]:} A lattice of increasing density, where the change in
  lattice density depends only on the r direction. In effect the
  original square lattice has been plastically deformed into the
  quadrilateral {\it p'q'r't'}. This diagram is geometrically
  incorrect as only the lines originally parallel to pq and sr should
  suffer a change of length, while those originally parallel to {\it
    ps} and {\it qr} should remain of the same length. {\it [C]:} The
  true deformed state of a lattice with increasing density, the
  lattice lines {\it pq} and {\it sr} are deformed into the circular
  arcs {\it p'q'} and {\it s'r'}. This state is geometrically correct
  as all of the lattice lines which were originally parallel to {\it
    ps} and {\it qr} are still of the same length. Note this
  construction only tells us how to calculate the curvature of lattice
  lines which were originally parallel to {\bf B}}
\label{bending}
\end{center}
\end{figure*}

The most remarkable feature of the numerical simulations is the
bending of lattice lines towards the edge of the cluster; which is
particularly strong in the larger systems - see for example Fig.
(\ref{very_large_systems}). This bending is due the fact that the
lattice contains an excess of dislocations of one sign. The original
argument explaining this phenomena was given by Nye \cite{nye}, who
showed that the curvature of the lattice is equal to the Burgers
vector density. However, Nye's exposition assumes that the lattice
spacing remains constant everywhere. Blindly applying Nye's result to
our system would predict that the lattice lines bend in the opposite
direction. With a slight modification we can adapt Nye's argument to
explain lattice curvature in crystals with a changing density.

Consider a square section of lattice {\it pqrs} of dimension $\Delta
r$ as shown in Fig. (\ref{bending}.A), the lattice has a constant
density everywhere and so there are an equal number of lattice lines
crossing each side of the square. If on the other hand the density is
increasing in the r direction it means there must be more lattice
lines crossing the side {\it sr} than {\it pq}. Drawing the Burgers
construction around the square leads to a closure failure.
Consequentially the square must contain an excess of edge dislocations
of one sign, the sum of their individual Burgers vector being equal to
the total Burgers vector {\bf B}. This situation is shown is Fig.
(\ref{bending}.B), it can be imagined that the total Burgers vector is
split into two equal parts and all the dislocations are contained
within the triangles {\it p't's'} and {\it q'u'r'} -- both of which
subtend an angle of $\theta/2$ -- it follows that
\begin{equation}
\frac{\theta}{2} \approx \frac{\frac{1}{2}nB}{\Delta r},
\label{eq:theta_over_two}
\end{equation}
where for small values of $\theta$ we can ignore higher order terms.
It is useful to imagine that the original lattice has been plastically
deformed into the quadrilateral {\it p'q's'r'}. However, this picture
is misleading as none of the lattice lines which were originally
parallel to {\it ps} and {\it qr} suffer a change in length; thus the
state shown in Fig. (\ref{bending}.B) is for illustrative purposes
only.  The true state of the deformed lattice is shown in Fig.
(\ref{bending}.C). By mapping the sides {\it pq} and {\it sr} of the
original square onto the circular arcs {\it p'q'} and {\it s'r'}, all
of the lattice lines originally parallel to {\it ps} and {\it qr}
remain of length $\Delta r$ -- at the expense of being no longer
parallel to each other. To find the curvature $k$ of the bending of
the arcs, let the length of the lines $o'p'$ and $o'q'$ be equal to L.
We use the relationship
\[
\Delta r=\frac{\theta}{k},
\]
where $k=1/L$. Upon substituting for $\theta$ from Eq.
($\!\!$~\ref{eq:theta_over_two}) and taking the continuum limit we
have
\begin{equation}
{\bf k}( {\bf r}  ) =\lim_{\Delta r\to\ 0} \frac{nB}{{(\Delta r)}^2}={\bf b}({\bf r}).
\label{eq:r_curve}
\end{equation}
Hence the curvature is equal to the Burgers vector density. If on the
other hand the lattice density is decreasing in the r direction then
we expect the sense of curvature to be reversed. Furthermore, for
lattice lines not parallel to the local Burgers vector, then the
curvature of the lattice line ${\bf k_o}$ depends on the angle
$\gamma$ it makes with the local Burgers vector \cite{nye}
\begin{equation}
{\bf k_o}({\bf r}) = {\bf k}({\bf r})\,\cos\gamma.
\label{eq:curvature_and_angle}
\end{equation}

\begin{figure}
\begin{center}
\includegraphics[width=0.6\columnwidth]{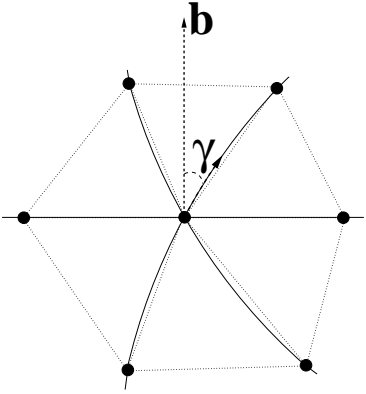}
\caption{ This diagram shows a hexagonal cell in a lattice with a
  non-zero Burgers vector density field {\bf b}. The hexagonal cell
  can be decomposed into three arcs which cross the center of the
  cell, if the arc makes an angle $\gamma$ with the vector {\bf b}
  then we expect its curvature to be given by $|{\bf b}|cos\gamma$.
  Thus as shown in this diagram the horizontal arc which is
  perpendicular to {\bf b} has no curvature.  }
\label{hexa_curve}
\end{center}
\end{figure}

\begin{figure}
\begin{center}
\includegraphics[width=0.6\columnwidth, angle=270]{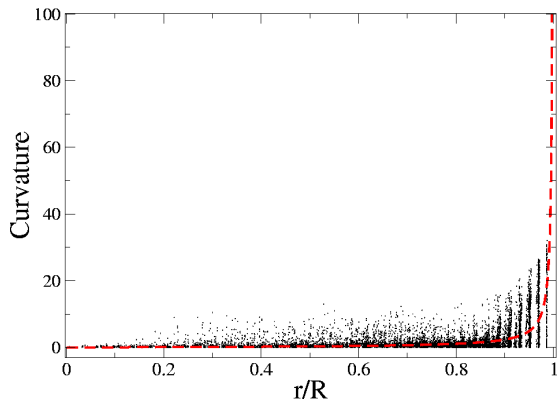}
\caption{Curvature of lattice lines in the system with 5000 charges in
  a hard wall confining potential. As discussed in the text, each
  hexagonal cell can be decomposed into a set of 3 arcs; for each such
  arc the curvature is computed and plotted against the radial
  distance of the parent hexagon -- this is shown by the black dots.
  Theoretically the maximum curvature that an arc can have is given by
  $|{\bf b}({\bf r})|$ which is shown by the red dotted line. Thus
  depending on the orientation of the lattice line, the curvature
  ought to range from 0 to $|{\bf b}({\bf r})|$. The fact that the
  lattice curvature exceeds $|{\bf b}({\bf r})|$ for some arcs is due
  to the fact that close to the core of a defect the curvature becomes
  much larger than that set by the continuum limit calculation.}
\label{curvature_of_lattice_lines}
\end{center}
\end{figure}

The next logical step is to show that the lattice curvature in the
systems which we have simulated is given by the Burgers vector
density. Consider the deformed hexagon shown in Fig.
(\ref{hexa_curve}) where the direction of {\bf b} is marked by an
arrow. We expect the curvature of the lattice lines to be described by
Eq.  ($\!\!$~\ref{eq:curvature_and_angle}). To make a connection with
our simulations, for the $N=5000$ system with a hard wall boundary
shown in Fig. (\ref{very_large_systems}), each hexagon is decomposed
into three arcs, each of which can be further decomposed into three
points as in Fig.  (\ref{hexa_curve}). By fitting the points to the
equation of a circle we determined the curvature of each arc. We
assume that each arc yields the curvature of the lattice at the center
of the hexagonal cell. Depending on the radial distance of the cell
from the center of the lattice we expect this curvature to have any
value between 0 and $|{\bf b}({\bf r})|$ (depending on the orientation
of the lattice line with respect to the local Burgers vector density
field). For the hard wall systems with 5000 charges, Fig.
(\ref{curvature_of_lattice_lines}) shows the curvature of each such
arc against radial distance; also plotted for comparison is the
Burgers vector density given by Eq. ($\!\!$~\ref{eq:mybvdensity}).
Ideally the curvature ought not to exceed the limit set by Eq.
($\!\!$~\ref{eq:mybvdensity}), however, this is not possible in
reality as lattice lines close to disclinations suffer much greater
curvatures than Eq. ($\!\!$~\ref{eq:mybvdensity}) would allow. From
the general similarity of the curvature data and Eq.
($\!\!$~\ref{eq:mybvdensity}) we conclude that the cause of the
bending of lattice lines is indeed due the plastic deformation of the
lattice.

Next we wish to make the connection between curvature as defined by
the work of Nye to some well-known results about parallel transport of
a unit vector about a disclination \cite{bowick3}. If a unit vector is
transported on a lattice along some path enclosing disclinations, then
when the vector returns to its original position its orientation will
have changed. The total change depends on the number of disclinations
enclosed and in a triangular lattice must be a multiple of $\pi/3$. To
show this connection we can invert Eq.  ($\!\!$~\ref{eq:curl_of_b}) to
express the density of the Burgers vector in terms of the disclination
charge enclosed
\begin{equation}
b_{\phi}
(r)
=
\frac{1}{r}
\int^{r}_{0}\!\,
\tilde{s}(r')
r'
d{ r'}
=
\frac{1}{2\pi r}
\int^{r}_{0}\!\,
\int^{2\pi}_{0}\!\,
\tilde{s}(r')
r'd{ r'}d{ \phi}.
\label{eq:b_and_s_densities}
\end{equation}

If as before we let $\Sigma (r)$ be the total disclination charge
within a disk of radius r, then using Eq. ($\!\!$~\ref{eq:r_curve}) we
can write Eq. ($\!\!$~\ref{eq:b_and_s_densities}) as
\begin{equation}
k_{\phi} (r)
=
\frac{\Sigma(r)}{2 \pi r}.
\label{eq:kurvature}
\end{equation}
This equation expresses the fact that the curvature at a distance r
from the center of the system depends on the total disclination charge
enclosed within a disk of radius r. For a given curve the curvature is
defined as the rate of change of the angle of its tangent vector
$\omega$, thus for the circular path enclosing a disclination charge
$\Sigma(r)$ we have
\[
\frac{1}{r}\frac{d{ \omega(r)}}{d { \phi}}=k_{\phi}(r).
\]
To find the total change in the angle of the tangent vector we
integrate over the length of the circular path giving,
\[
\omega(r)=\int^{2 \pi}_{0}\!\, rk_{\phi}(r) d{ \phi}=\Sigma(r),
\]
where in a real lattice the disclination charge is quantized.  This
result explains why the orientation of the lattice cells in images
such as Fig. (\ref{very_large_systems}) is observed to rotate upon
traversing a circular path centered on the origin.

It should be noted that as long ago as 1955 the geometry of imperfect
lattices had been developed by Kondo and co-workers into
non-Riemann differential geometry \cite{kondo}. By analogy to
general relativity, one can think of an undeformed lattice as a region
of flat space, which becomes warped in the presence of defects. The
authors demonstrated that dislocations and disclinations generate
torsion and curvature respectively. 

\subsection{Conformal Crystals}

\begin{figure*}[htp]
  \begin{center}
    $\begin{array}{c@{\hspace{3.0cm}}c}
      \epsfxsize=6cm
      \epsffile{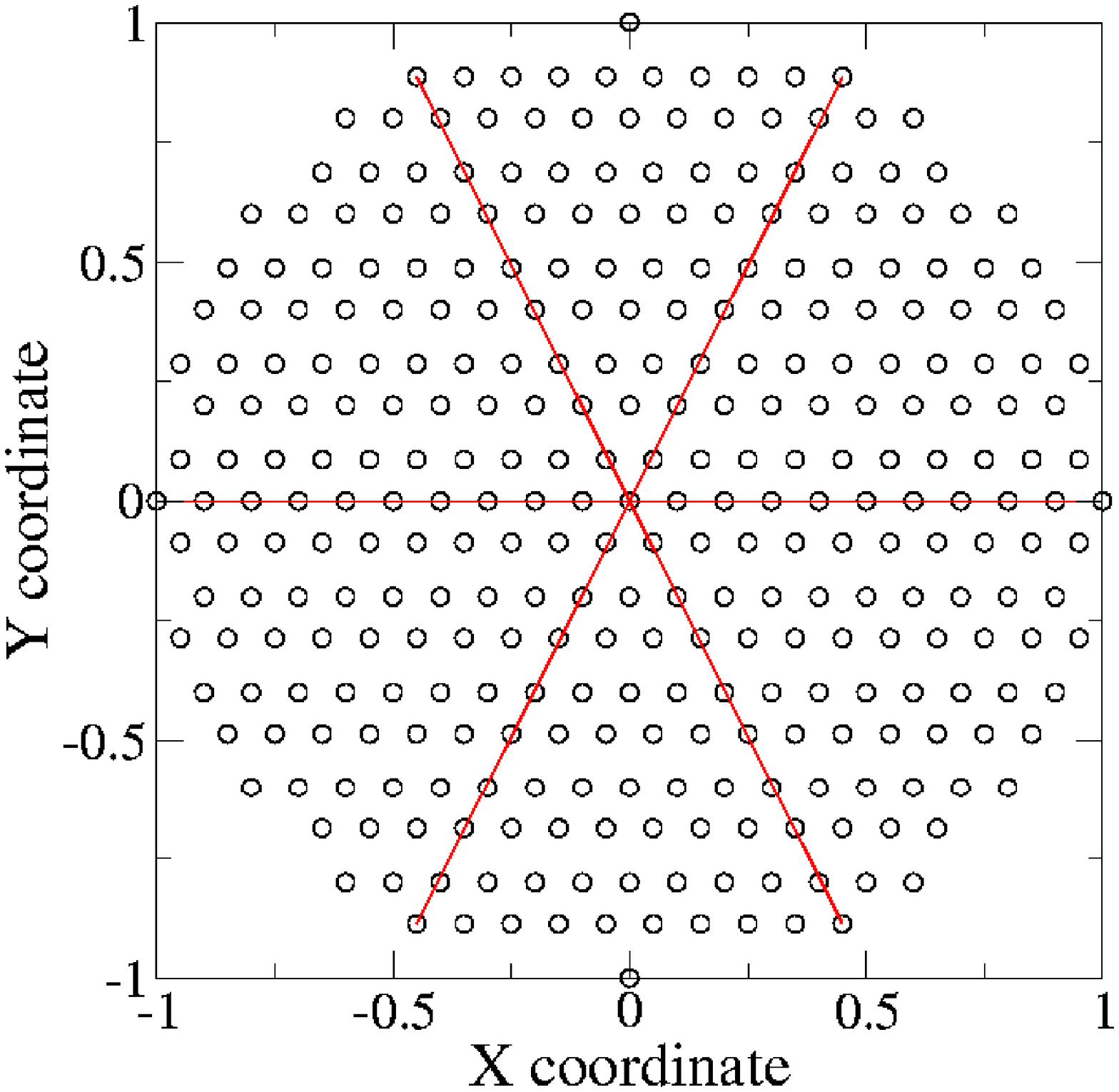} &
      \epsfxsize=6cm
      \epsffile{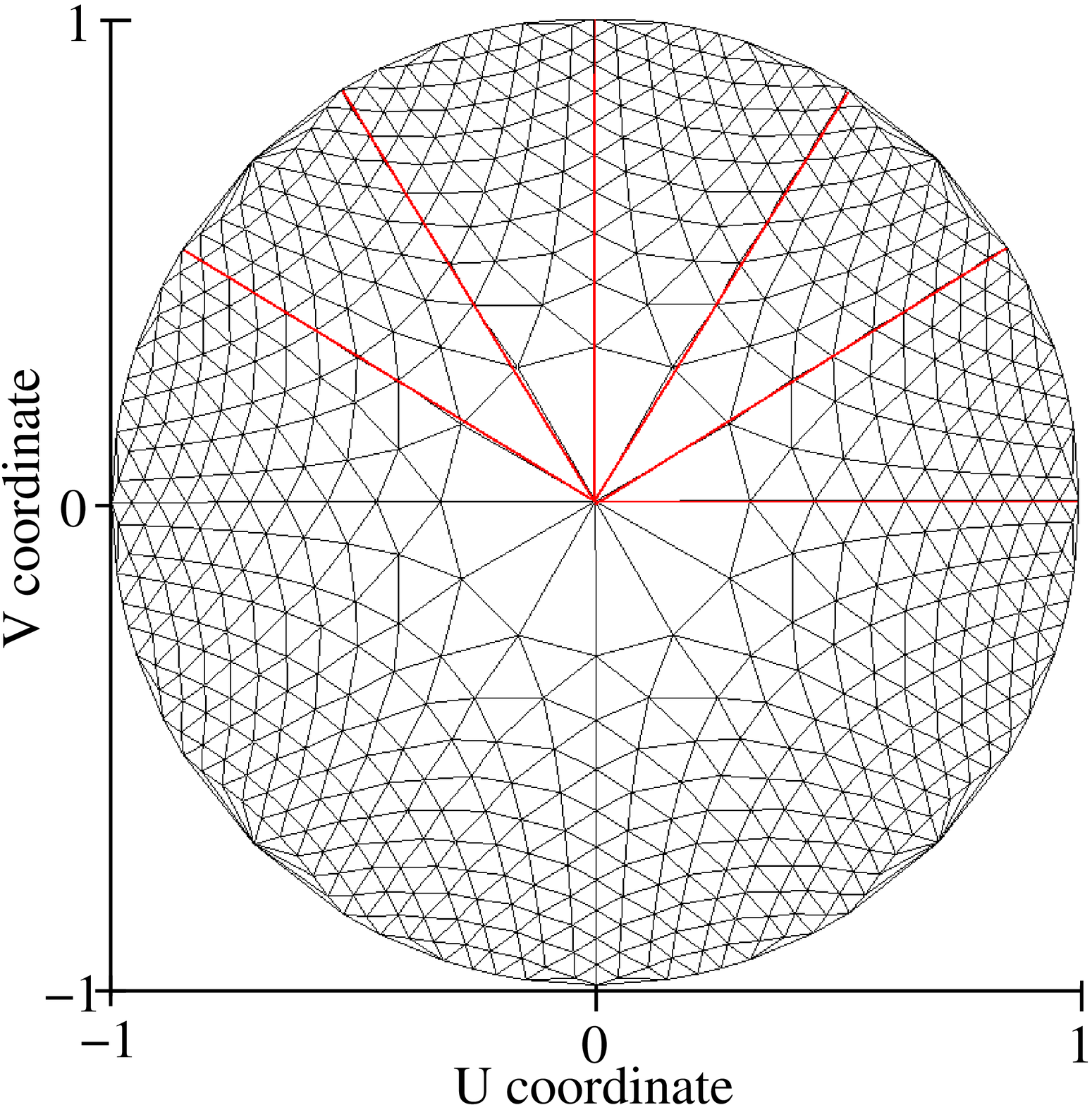} \\ [0.3cm]
    \end{array}$
  \end{center}
  \caption{{\bf Left:} a circular cut out of a triangular lattice in
    the z plane, where the angle between each successive red line is
    $\pi/3$. {\bf Right:} the effect of the transformation
    $w=z^{\frac{1}{2}}$ in the image plane, where the lattice has been
    produced by allowing the range of the z plane to extend to
    $\phi<4\pi$. Conformal transformations are locally angle
    preserving transformations, therefore an infinitesimal hexagonal
    lattice cell in the z plane is not deformed upon mapping to the
    image plane.  However, this angle preserving property of the
    transformation breaks down at certain points which are known as
    critical points.  For the transformation $w=z^{\frac{1}{2}}$ this
    occurs at the origin, where instead of preserving the angle
    $\pi/3$, it is halved. Note that the lattice cells are more
    distorted closer towards the critical point.}
  \label{conformal_transformation}
\end{figure*}

The systems we have discussed thus far have been constructed using
optimization algorithms to generate ground state
configurations. Surprisingly there exists an unusual class of 2D
lattices which have a non-uniform density but which can be constructed
by a purely analytical method. These structures are known as conformal
crystals \cite{rothen1}. As the name suggests the positions of the
lattice sites can be obtained in the image plane by applying a
conformal transformation to a regular lattice in the z-plane.

An example of a conformal lattice with circular symmetry is shown in
Fig. (\ref{conformal_transformation}). As evident from the way in
which the lattice lines curve towards the edge and the increasing
density, there is a resemblance between the conformal crystal and the
clusters studied in Section IV. Initially it was hoped that these
conformal crystals might help explain the origin of the lattice
curvature observed in the other 2D systems, but this did not turn out
to be the case. In this section we show that conformal crystal can be
regarded as a giant disclination.

Originally the idea of conformal crystals was used to describe a
structure formed by a cluster of mutually repelling magnetized spheres
dubbed ``gravity's rainbow''. In an experiment, metallic spheres were
 confined to a thin rectangular box which is placed in a magnetic
field, the field induces a magnetic moment in each of the spheres
which causes them to repel. Under the action of gravity the spheres
crystallize into a lattice with non-uniform density, which consisted of
a series of arch-like structures. The authors suggested that the
unusual lattice could be obtained by a conformal transformation of a
regular triangular lattice \cite{rothen2}. 

Consider for example a regular triangular lattice in the z-plane such
as that shown in Fig. (\ref{conformal_transformation} - left). By
applying an analytical transformation $w=f(z)$ the corresponding
coordinates in the w-plane are
\[
u+iv=f(x+iy),
\]
where $w=u+iv=re^{i \theta}$ and $z=x+iy=r'e^{i \phi}$. In the case of
$f(z)=z^\frac{1}{2}$ the result of the transformation is shown in Fig.
(\ref{conformal_transformation}). This transformation belongs
to a set of transformations
\begin{equation}
w=Cz^{\frac{1}{\chi}}\,\,\,{\rm or}\,\,\,w=e^{Cz},
\label{eq:radial_symm}
\end{equation}
which yield conformal lattices of circular symmetry. Here we shall
only concern ourselves with the first of these transformations.

Conformal transformations have three important features. Firstly they
are locally angle preserving (isogonal) transformations. This means
that upon mapping an infinitesimal hexagonal lattice cell in the z
plane to the image plane, the cell will still have all its internal
angles equal to $\pi/3$. It is important to note that because of the
local nature of the angle preserving property this is only strictly
true for an infinitesimal lattice cell. For a real lattice with a
well-defined lattice spacing, the mapping distorts the shape of the
hexagonal cells. This distortion is stronger towards the center of the
lattice than the edge, see Fig. (\ref{conformal_transformation}) for
an illustration of this. At the origin the conformality of the
transformation breaks down completely.  In the case of the
transformation $w(z)=z^{1/2}$, instead of preserving angles, the angle
of $\pi/3$ at the origin is halved in the image plane. No matter how
small the lattice cell enclosing the center in the z-plane is made,
this breakdown of conformality will remain. If the conformality of an
otherwise conformal mapping breaks down at a particular point, then
that point is called a critical point of the mapping. The critical
points of a conformal transformation exist at any point where either
$|dw/dz|$ or its inverse is equal to zero \cite{jeffreys}. Secondly, a
lattice with constant density $\rho_z$ in the z-plane will have a
density \cite{rothen1}
\begin{equation}
\rho_w=\rho_z\left|\frac{dw}{dz}\right|^{-2},
\label{eq:conformal_density}
\end{equation}
in the w-plane. Thirdly, it was shown that if the transformation is
conformal then the density of lattice points in the image plane is
constrained by the following condition \cite{rothen2}
\begin{equation}
\nabla ^{2}\ln \rho(r)=0, 
\label{eq:conformal_cond}
\end{equation}
which interestingly is also the condition that the lattice has no
disclination charge induced by a varying density, see Eq.
($\!\!$~\ref{eq:curl_of_b}).

\begin{figure}[t]
\begin{center}
\includegraphics[width=0.7\columnwidth]{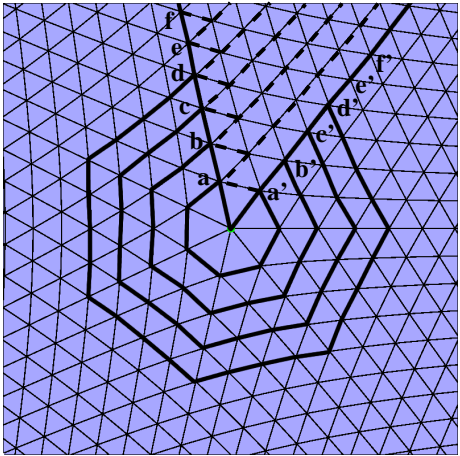}
\caption{A conformal lattice generated by the transformation
  $w=z^{6/7}$ which is similar to a disclination with charge -1. Also
  shown are a series of Burgers circuits enclosing the central point.}
\label{7pt_conformal_lattice}
\end{center}
\end{figure}

Thus a circularly symmetric conformal crystal such as the one shown in
Fig. (\ref{conformal_transformation}) is locally hexagonal but has
curvature. It has an increasing density and except for the point
at the center has no apparent internal defects (by which we mean that
a triangulation only shows internal points which have six nearest
neighbors). Yet the framework developed in Sections 2.3 and 2.4
demonstrates that a change in the lattice density must be accompanied
by lattice defects, which in turn are responsible for curvature. How
are these two seemingly conflicting statements to be resolved?  A clue
is provide by the point at the center of the lattice with the
anomalous coordination number.

Consider the conformal lattice generated by the transformation
\[
w(z)=z^\frac{6}{7},
\]
which is shown in Fig. (\ref{7pt_conformal_lattice}). Superimposed on
top of the conformal lattice are a number of Burgers circuits which
{\it enclose} the central point. Whereas previously we were dealing
with a continuum, in which the defects were assumed to form a gas
throughout the system, here there is only a single defect at the
center of the system. Since there are no other defects, any Burgers
circuit which does not enclose this central point will close. As shown
the circuits start at a,b... and end at a',b'..., in each case there
is a closure failure, which increases with distance from the central
point; indeed the successive Burgers circuits trace out a wedge. Thus
we propose the central point in a conformal lattice can be thought of
as a ``disclination''. The same rules which apply to a disclination
also apply here, i.e. the wedge angle is quantized by the symmetry of
the lattice and the wedge in turn can be decomposed into a series of
half planes \cite{friedel}. To see how a wedge can be decomposed into
a series of half planes, note that the difference in the closure
failure between any two successive Burgers circuits, such as aa' and
bb', is always one lattice spacing, which implies that between a pair
of circuits an extra half plane has been inserted at the points
labeled a,b... The question of where the half planes have been
inserted is arbitrary as it depends on where the Burgers circuits
start and end. Thus as for a disclination the relationship given by
Eq. ($\!\!$~\ref{eq:closure_faliure_around_a_disclination}) also holds
for a conformal lattice. In the case of $w(z)=z^\frac{6}{7}$ (also for
a -1 disclination) the wedge angle is given by $\Omega=\pi/3$. By
considering a series of circular concentric circuits whose origin
coincides with the critical point of the transformation, it is
possible to draw a series of Burgers circuits and calculate the
Burgers vector density using the approach outlined in section III.B,
in either case the result is still given by Eq.
($\!\!$~\ref{eq:bvdensity}).

Since at the critical point the lattice ceases to be conformal, Eq.
($\!\!$~\ref{eq:conformal_cond}) is true everywhere except for the
origin. This suggests that it can
be written as
\begin{equation}
\nabla ^{2}\ln \rho(r)=\nu \delta^{(2)} ({\bf r}), 
\label{eq:conformal_cond_new}
\end{equation}
where $\nu$ is an undetermined constant and $\delta^{(2)}$ is the two
dimensional delta function. To find $\nu$, we are going to assume that
the point at the center of the lattice is a disclination with charge
$\tilde{s}(r)$. We have the following relationship between Eq.
($\!\!$~\ref{eq:conformal_cond_new}) and Eq.
($\!\!$~\ref{eq:curl_of_b})
\begin{equation}
2\tilde{s}(r)=\nabla ^{2}\ln \rho(r)=\nu \delta^{(2)}({\bf r}).
\label{eq:conf_and_disc}
\end{equation} 
Substituting $w(z)=Cz^{\frac{1}{\chi}}$ into
Eq. ($\!\!$~\ref{eq:conformal_density}) gives
\begin{equation}
\rho(r)=
\rho_z
\chi^2 
r'^{\frac{2(\chi-1)}{\chi}}
=\rho_z\chi^2r^{2(\chi-1)},
\label{eq:conden}
\end{equation}
where we have used the relationship $r=r'^{\frac{1}{\chi}}$. Thus we can
write Eq. ($\!\!$~\ref{eq:conf_and_disc}) as 
\begin{equation}
2\tilde{s}(r)=2(\chi-1)\nabla ^{2}\ln r=\nu \delta^{(2)} ({\bf r}),
\label{eq:wowgreen}
\end{equation}
Recognizing $\nabla ^{2}\ln r= 2\pi\delta^{(2)} ({\bf r})$ as the two
dimensional Green's function and canceling out the factor of 2,
Eq. ($\!\!$~\ref{eq:wowgreen}) becomes
\[
\tilde{s}(r)=2\pi(\chi-1)\delta^{(2)}({\bf r})=\nu'\delta^{(2)}({\bf r}),
\]
thus the central point of the conformal transformation has a
disclination charge density of $\nu'=\nu/2=2\pi(\chi-1)$.  The total
disclination charge contained within the disk can be found by simply
integrating over its area, thus
\begin{eqnarray}
\Sigma(r)
&=&
\int^{R}_{0}\!\,\int^{2\pi}_{0}\!\,\tilde{s}(r)rdrd\theta
\nonumber
\\
&=&
2\pi(\chi-1)\int^{R}_{0}\!\,\int^{2\pi}_{0}\!\,\delta^{(2)}({\bf r})rdrd\theta
\nonumber
\\
&=&
2\pi(\chi-1),
\end{eqnarray}
In the case of $\chi=2$ which corresponds to the transformation
$w(z)=z^{1/2}$, the central point of the transformation has a
disclination charge of $6(\pi/3)=2\pi$, i.e. a disclination consisting
of 6 wedges, each of which subtend an angle of $\pi/3$ in the z
plane.

This realization that the central point in a conformal lattice is
actually a disclination can be used to calculate the lattice
curvature. The conformal lattice shown in Fig.
(\ref{conformal_transformation}) was generated by applying the
transformation $w=z^\frac{1}{2}$ to a hexagonal lattice, the original
lattice contained $N_{z}$ points within a disk of radius R, thus
$\rho_{z}=N_{z}/{\pi R^2}$, setting $\chi=2$ in Eq.
($\!\!$~\ref{eq:conden}) yields
\[
\rho_{w}(r)=
4r^{2}\rho_{z},
\]

Using the framework developed in Section 2.4 we know that the maximum
curvature $k$ is related to the density of the Burgers vector by
Eq. ($\!\!$~\ref{eq:r_curve}), thus
\begin{equation}
k (r)
=
|{\bf b}({\bf r})|
=
\frac{1}{2}
\frac{d}{dr}
\ln \rho_w(r)
=
\frac{1}{2}
\frac{d}{dr}
\ln 4r^2\rho_z
=
\frac{1}{r},
\label{eq:conf_curve}
\end{equation}
where we have used Eq. ($\!\!$~\ref{eq:bvdensity}). On the other hand,
using Eq. ($\!\!$~\ref{eq:closure_faliure_around_a_disclination}) with
$\Omega=2\pi$ gives $|{\bf B}(r)|=(2\pi)r$. Inverting the relationship
given by Eq. ($\!\!$~\ref{eq:curl_of_b}) yields
\begin{equation}
k (r)
=
|{\bf b}(r)|
=
\frac{1}{2\pi r}
\frac{d}{dr}
|{\bf B}(r)|
=
\frac{1}{2\pi r}
\frac{d}{dr}
2\pi r
=
\frac{1}{r}
\label{eq:sloop}
\end{equation}

\begin{figure}[t]
\begin{center}
\includegraphics[width=0.6\columnwidth , angle=270]{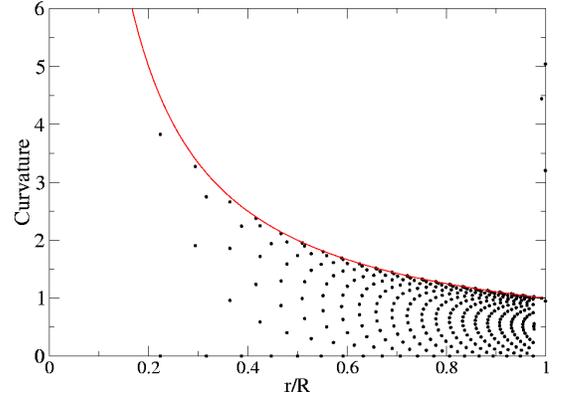}
\caption{ 
Curvature of the
lattice lines for a conformal crystal generated by the transformation
$w(z)=z^\frac{1}{2}$. The red line gives the maximum curvature using
Eq. ($\!\!$~\ref{eq:conf_curve}). The black dots give the actual
curvature of the lattice line, as calculated by the method outlined
in Section 4.5.
}
\label{conformal_lattice_curvature}
\end{center}
\end{figure}

Alternatively, using Eq. ($\!\!$~\ref{eq:kurvature}), we can assume
that the disclination charge enclosed is equal to $2\pi$ and this
gives the same result. Thus a comparison can be made between this
analytical result and the actual measured lattice curvature (just like
we did for $N=5000$ in the hard wall case),
Fig. (\ref{conformal_lattice_curvature}). There is perfect agreement
between the two curves which leads to the conclusion that a conformal
lattice can be thought of as a type of disclination.

\subsection{Experimental Realizations}

The system with the hard wall confinement could potentially be
realized in a number of different physical contexts. One possible
experimental situation involves a collection of polystyrene beads on a
disk, located on the edge of which is an insulated boundary. The beads
are then exposed to ionizing radiation and the liberated electrons are
sucked out of the system by an electrode, leaving behind a collection
of positively charged beads. The cluster can then be annealed to the
ground state by shaking the disk.

Alternatively the system could be realized using colloidal particles.
For these systems one would have to ensure that the Yukawa screening
length was made larger than the system size. By using less polar
organic solvents, screening lengths as long as $12 \mu m$ can be
achieved \cite{yethiraj}. Thus to realize large systems one would need
colloidal particles with diameters of, say, $0.1 \mu m$.

The system with parabolic confinement is of considerable interest to
the field of quantum dots. For review articles discussing
the fabrication of quantum dots and the harmonic confining
approximation, see \cite{alhassid} and \cite{reimann} respectively and
the references contained.

\subsection{Other Systems}

There are a number of systems in which lattice curvature is quite
prominent; one such system is the growth of crystals in amorphous
films \cite{kolosov}. Up until now the suggestion has been that these
systems possess a conformal geometry because the lattice lines are
curved. This work demonstrates that this is not necessarily so. The
condition that a system has a conformal geometry is very strict, i.e.
the lattice density has to obey Eq. ($\!\!$~\ref{eq:conden}). It is
possible that these systems contain an excess of disclination charge
in the interior, which in turn results in the bending of lattice
lines. It would be interesting to re-examine such systems in light of
the results of this paper.  Even the original experiment which sparked
the interest in conformal crystals, the so-called gravity's rainbow
structure, has been shown not to be stable \cite{klos}, meaning that
it is likely that the system does not form a perfect conformal
crystal. Numerical simulations suggest that the system is composed of
domains which are separated by defects \cite{klos}; these defects may
be the actual cause of the lattice curvature in this system. Other
interesting systems in which almost perfect conformal crystals have
been generated include ferrofluid foams in magnetic fields
\cite{elias} and soap foams \cite{weaire}; in some cases the resulting
structure contains internal defects. It is hoped that the present work
may give some insight into their role.

\acknowledgements One of us (AM) would like to thank EPSRC for
financial support. We would also like to thank Matthew Hastings and
Paul McClarty for useful discussions.


\addcontentsline{toc}{chapter}{References}
\bibliographystyle{nonspacebib}
\bibliography{my_ref}

\end{document}